\def\VEL{\:{\rm km\:s^{-1}}}

\documentclass[12pt,preprint]{aastex}

\begin{document}

% Additional private definitions that appear to work only inside document

\newcommand{\MSOL}{\mbox{$\:M_{\sun}$}}

\newcommand{\EXPN}[2]{\mbox{$#1\times 10^{#2}$}}
\newcommand{\EXPU}[3]{\mbox{\rm $#1 \times 10^{#2} \rm\:#3$}}  % exponent with units
\newcommand{\POW}[2]{\mbox{$\rm10^{#1}\rm\:#2$}}

% End of defining things

\title{\bf{Modeling the Spectral Signatures of Accretion Disk Winds:
A New Monte Carlo Approach}}

\author{Knox S. Long}

\affil{Space Telescope Science Institute, \\ 3700 San Martin
Drive, \\ Baltimore, MD 21218}

\author{Christian Knigge}

\affil{Department of Physics \& Astronomy,\\
University of Southampton, \\
Southampton SO17 1BJ
UK}

\begin{abstract}
Bipolar outflows are present in many disk-accreting astrophysical
systems. In disk-accreting cataclysmic variables (CVs), these
outflows are responsible for most of the strong features in the
ultraviolet spectra of these systems. However, there have been few
attempts to model these features quantitatively. Here, we describe
a new, hybrid Monte Carlo/Sobolev code, which allows us to
synthesize the complete spectrum of a disk-dominated, mass-losing
CV.

The line profiles we calculate for C IV resemble those calculated
by previous workers when an identical geometry is assumed.
However, our synthetic spectra exhibit not only the well-known
resonance lines of O VI, N V, Si IV and C IV, but, with an
appropriate choice of mass-loss rate and wind geometry, also many
lines originating from excited lower states. Many of these lines
have already been seen in the far ultraviolet spectra of CVs
obtained with HUT, FUSE, and HST. In order to illustrate the
degree to which we are currently able to reproduce observed
spectra, we finally present a preliminary fit to the Hopkins
Ultraviolet Telescope spectrum of the dwarf nova Z Cam in
outburst.

\end{abstract}


\section{Introduction}

Non-magnetic cataclysmic variables stars (CVs) are semi-detached
binary systems in which a late-type companion star loses mass to a
white dwarf (WD) primary via Roche-lobe overflow onto an accretion
disk. The discovery of winds in high-state CVs -- from the
presence of blue-shifted absorption troughs and P-Cygni profiles
in the ultraviolet (UV) resonance lines -- dates back about two
decades \citep{heap1978,cordova1982}. However, most of the
fundamental {\em theoretical} questions concerning the nature and
origin of this mass loss are only now beginning to be addressed
\citep{drew2000}.

Recent work suggests that the impact of the outflowing material on
{\em observations} of CVs may have been underappreciated, in that
its presence may leave more numerous and subtle observational
signatures than just the shapes of the ``classical'' UV wind
lines. For example, it has been suggested that a powerful disk
wind may be responsible for the peculiar behavior of the SW~Sex
stars \citep{honeycutt1986,dhillon1995}, and that the often
single-peaked optical emission line profiles seen in these systems
(and also in other nova-like variables [NLs]) are the direct
consequence of a wind-induced velocity gradient in the
line-emitting material \citep{murray1996}. Similarly,
\cite{knigge1997} suggested that virtually all of the lines in the
UV spectrum of a typical high-state CV, Z Cam, are formed in the
outflow, either in the supersonic portion of the wind or in a
lower velocity portion of the wind near the interface with the
disk photosphere. Based on an analysis of low resolution
(1140-8950 \AA) spectra of the eclipsing novalike variable UX~UMa,
\cite{knigge1998} then suggested that continuum emission from an
optically thin disk wind and/or atmosphere provided one possible
explanation for the absence of a clear Balmer jump and the flatter
than expected  UV continua of high-state CVs
\citep{wade1984,wade1988}. Optical spectra of at least one CV, BZ
Cam, shows P-Cygni features in the hydrogen and helium lines that
are most easily interpreted in terms of a highly variable wind
\citep{ringwald1998}.  In addition, the EUVE spectrum of the dwarf
nova (DN) U~Gem in outburst appears to be dominated by strong wind
features to such an extent that any attempt at a detailed spectral
analysis {\em must} account for the outflowing material
\citep{long1996}. Finally, there is also direct evidence that CV
winds act as partially ionized absorbers of (soft) X-rays in
high-state, non-magnetic systems \citep{baskill2001}.

The first attempts to interpret UV line profiles in high-state
disk-dominated CVs were quite naturally based on wind modeling
theory that was being developed to interpret the UV spectra of
early-type stars \citep{mauche1987, drew1987, vitello1988}. These
studies, along with a burgeoning amount of observational analysis,
indicated the mass loss rates in CV winds are a substantial
fraction ($\sim$10\%) of the accretion rate, that the acceleration
length scale is substantial (20-100 $R_{wd}$), and that the winds
are strongly affected by rotation, presumably reflecting an origin
in the rapidly rotating inner disk.

The evident importance of rotation led to a second generation of
radiative transfer codes that provided for azimuthally symmetric
flows. In the CV wind code described in \cite{SV1993}, hereafter
SV93, the wind ionization structure is computed first and the
radiative transfer then calculated using the Sobolev (SA) and
disconnected approximations. This method is comparatively fast,
but does not treat multiple scattering correctly in detail. It was
applied by \cite{vitello1993} to show that, depending on the wind
geometry, wind mass loss rates of order 1-10\% of the accretion
rate could account for the C IV line profiles of RW Sex, RW Tri
and V Sge . By contrast, the Monte Carlo (MC) code described by
\cite{knigge1995}, hereafter KWD95, simulates the transfer of
resonance line photons through an outflow exactly, but does not
perform an ionization calculation. It was used by
\cite{knigge1997b} to model high resolution time-resolved spectra
of the C IV region in UX~UMa to show that a relatively dense,
slowly outflowing transition region between the disk photosphere
and the fast wind was necessary to understand the profiles in that
system.

Both of these radiative transfer codes have limitations. Neither
method solves self-consistently for the thermal balance in the
wind (constant wind temperature is assumed); neither is designed
to allow for the thermal creation or destruction of line photons
in the wind  \citep[but see][for an approximate treatment in the
MC context]{knigge1997}. Finally, both of these codes are intended
for modeling of individual spectral lines; as a result it is
usually unclear whether a wind model intended to reproduce the C
IV profile has any predictive power for other wind signatures.

Here, we present a new, hybrid SA/MC method designed to provide
improved spectral modeling of azimuthally symmetric winds. Various
aspects of our method build on the work of Lucy and collaborators
\citep{mazzali1993,lucy1999a,lucy1999b} who have to-date
considered only spherically symmetric situations. In addition to
combining the respective strengths of the SA and MC methods in the
(conservative) transfer of resonance lines, our method solves
self-consistently for the thermal and ionization balance in the
outflow and accounts for the destruction and creation of both line
and continuum photons in the wind. As a result, our method can be
used to synthesize any part of the emergent, wind-affected
spectrum, allowing detailed analyses of previously neglected wind
signatures.

\section{Method}

\subsection{Basics}

The idea behind our Monte Carlo algorithm is simple: a large
number of suitably randomized and distributed energy packets
(``photons'') are generated to represent the radiation field of
the accretion disk, the WD, the boundary layer (BL) and the wind
itself. These photons are followed as they make their way through
the wind, where they can be scattered by resonance lines,
scattered by electrons, absorbed in photoionization events or
destroyed via collisional de-excitations. After accounting for
these interactions,
 all photons escaping towards the
observer are binned in wavelength to yield the wind-attenuated
spectrum of the system as viewed from the observer's viewing
angle.

Not surprisingly, the {\em implementation} of this simple idea is
more complex. Each complete run of our code actually consists of
two separate sets of Monte Carlo simulations. In the first, the
ionization and temperature structure of the outflow is calculated
by iteration. To this end, an initial wind structure is assumed
first, and a family of photons representing the entire system
luminosity is generated and followed through the outflow. Photons
scatter when the scattering optical depth
\begin{equation}
\tau_{scat}\ge -ln(1-p) \label{scat_opt_depth}
\end{equation}
where p is a random number between 0 and 1 generated with a
uniform distribution. The characteristics of all photons that pass
through each grid cell are stored and used to obtain better
estimates for the ionization state and temperature of that cell. A
new iteration is then started by generating a new family of
photons. After a number of iterations, the temperature and
ionization structure of the outflow converges. This converged
structure is used as a fixed input to the second set of MC
simulations. In this set, the detailed emergent spectrum is
calculated for a user-defined wavelength range. This can be as
narrow as a single line profile or as wide as the entire spectral
energy distribution. The wind structure is not (and, generally
speaking, cannot be) refined any further at this stage, since only
photons in a restricted frequency range are generated.

In the following sections, we will provide a more detailed look at
the inner workings of our code. We will begin with a description
of the two kinematic wind models we have implemented to date.

\subsection{Wind Models}
The underlying geometry of our Monte Carlo code is cylindrical.
The center of the WD defines the origin of the grid. The code is
designed to handle winds that are symmetric with respect to
reflection through the disk plane, as well as azimuthally
symmetric with respect to the disk axis. As part of the
initialization process, a grid of points with logarithmic (or
linear) spacing is established, onto which the model of the wind
is impressed.   Linear interpolation is used to assure that
variables, e.g. velocity and velocity gradients, are continuous
throughout the wind. Provision is made for defining regions in
which the wind exists and regions in which it does not. These
features make it straightforward to support various wind model
geometries. Despite the fact that the grid is cylindrical, one can
accommodate models in which the wind occupies a biconical region
emerging at arbitrary angles from the disk, or alternatively one
that is approximately spherical. In the discussion that follows,
we will use terminology specific to CVs, but the code is not
limited to that astrophysical situation.

We currently implement the two kinematic descriptions of CV winds
used by SV93 and by KWD95 in our code. We selected these because
they are the two prescriptions that have been used most
extensively to model C IV line profiles.  Both are biconical wind
flows, and both assume that material flows out from the disk along
streamlines. Both separate the velocity into poloidal ($v_l =
\sqrt{v_{\rho}^{2}+ v_z^{2}}$) and azimuthal components
$v_{\phi}$. Both assume that the outflowing material shares the
Keplerian speed of the accretion flow at the footpoint of a
streamline and that the material in the wind conserves specific
angular momentum on its way out of the system.  As a result the
azimuthal velocity is large close to the disk, and is largest at
the inner edge of the disk where the Keplerian velocity of the
disk is greatest.

In the SV93 prescription (see Fig.\ \ref{sv_cartoon}), the wind
emerges from the disk between $r_{min}$ and $r_{max}$ along
streamlines whose orientation with respect to the system are
described by an angle\footnote{We note in passing that in a few
cases we have consolidated for clarity within this article the
notation of the two models.  We trust that this will not be
confusing for those who refer back to SV93 and KWD95}
\begin{equation}
\theta = \theta_{min} + (\theta_{max} - \theta_{min}) x^{\gamma}
\end{equation}
where
\begin{equation}
x=\frac{r_o-r_{min}}{r_{max}-r_{min}}
\end{equation}
and $r_o$ refers to the position of the footpoint of a streamline.
A highly collimated flow will result if the difference between
$\theta_{min}$ and $\theta_{max}$ is small while $\gamma$ controls
the concentration of the stream lines toward either of the two
wind boundaries.

The poloidal velocity along the streamlines is defined to be
\begin{equation}
v_l=v_o + (v_{\infty}(r_o) - v_o)\left(
\frac{(l/R_v)^{\alpha}}{(l/R_v)^{\alpha}+1} \right)
\end{equation}
The scale length $R_v$ and the exponent $\alpha$ control the
acceleration of the wind between a fixed  velocity $v_o$, normally
6 $\VEL$, at the base of the wind and the terminal velocity
$v_{\infty}(r_o)$. The terminal velocity of each streamline varies
depending on the location of the streamline in the inner and outer
disk, being characterized as a fixed multiple of the escape
velocity at the footpoint of the streamline. Thus the poloidal
velocity is greatest for stream lines that originate from the
inner regions of the disk, since the gravitational potential that
must be overcome is greatest there.

The mass loss per unit surface area $\delta \dot{m}/\delta A$ of the disk is
controlled by a parameter $\lambda$ such that
\begin{equation}
\frac{\delta\dot{m}}{\delta A} \propto \dot{m}_{wind} r_o^{\lambda} cos(\theta(r_o))
\end{equation}
With this prescription, the overall mass loss rate declines with
radius if $\lambda$ is somewhat less than -2.

To use the SV93 prescription, therefore, one must provide the
basic parameters of the system, the mass of the WD, the accretion
rate, the inner and outer radius of the disk, and in addition, for
the wind $\dot{m}_{wind}$, $r_{min}$, $r_{max}$, $\theta_{min}$,
$\theta_{max}$, $\gamma$ ,$R_{\nu}$, $\alpha$, $\lambda$, and the
multiple of the escape velocity to be used for $v_{\infty}$.

The KWD95 characterization of the wind has somewhat fewer
parameters. In it, as indicated in Fig.\ \ref{knigge_cartoon}, the
outflow in each hemisphere of the disk system is described as if
it originated from a point a distance D on the other side of the
disk plane. The inner and outer boundary of the wind is determined
by the WD radius and the outer radius of the disk. For large D the
wind will be highly collimated along the axis of the disk, and for
very small D it will be highly collimated in a thin region with
large opening angle. The least collimation occurs when
\begin{equation}
D=\sqrt{R_{disk} R_{wd}}.
\end{equation}
For example, if $R_{disk}=50 R_{wd}$, then the wind with the least
collimation angle has a value of D of $7.07 R_{wd}$ and extends
from $\theta_{min}$ of 8\degr\ to $\theta_{max}$ of 82\degr. On
the other hand, if D were 10$\times$ larger, then $\theta_{min}$
would be 0.8\degr\ and $\theta_{max}$ would be
35\degr.\footnote{Rather than use D, the distance along the z axis
from which streamlines originate, KWD95 characterized their models
in terms of the ratio $D/D_{min}$ where $D_{min}$ corresponds to
the point at which the solid angle of the disk is maximized from
an observer located at $D_{min}$.  This is approximately
$0.5\sqrt{R_{disk} R_{wd}}$, with the exact value of the
coefficient depending on the ratio of $R_{disk}$:$R_{wd}$. We have
chosen to deprecate the use of $D/D_{min}$, except when it is
necessary to refer back to models calculated by KWD95.}

In their description, the mass loss rate per unit surface area
\begin{equation}
\frac{\delta \dot{m}}{\delta A} \propto T(R)^{4\alpha}
\end{equation}
For $\alpha = 0$, mass loss rate per unit area is uniform, while
for $\alpha = 1$ mass loss is proportional to the luminous flux of
the disk.

Like SV93, KWD95 adopt a velocity law reminiscent of those used to
describe stellar winds, but modified slightly to reflect the fact
the wind is arising from the disk. More specifically, they define
the poloidal velocity
\begin{equation}
v_l=c_{s} + (v_{\infty} - c_{s})\left(1- \frac{R_{v}}{l+R_{v}}
\right)^{\beta}
\label{knigge_vel}
\end{equation}
where $c_s$ is the speed at the base of the flow, $R_v$ is the
scale length for velocity law,  and $\beta$ is the exponent that
determines the number of scale lengths over which the wind
accelerates. Like SV93, $v_{\infty}$ is defined as a multiple of
the escape velocity at the footpoint of each stream line.

KWD95 assume the poloidal velocity at the base of the wind is just
the sound speed of the disk surface, defined as
\begin{equation}
c_s(R) = 10 \sqrt{\frac{T_{eff}(R)}{10^4 K}} \VEL.
\end{equation}
where $T_{eff}$ is given by equation \ref{eq-disk} below
\citep{frank1985}.

Thus for the KWD95 prescription, one must define the basic
parameters of the system as was the case for SV93, and then
$\dot{m}_{wind}$, D, $\alpha$, $R_s$, $\beta$, and the multiple of
the escape velocity to be used for $v_{\infty}$. The KWD95
prescription has fewer parameters, mostly due to a less flexible
prescription for controlling the direction of streamlines and for
limiting the region on the disk from which the wind emerges.
Depending on one's perspective, the differences in flexibility
between the two prescriptions can be taken as an advantage or a
disadvantage. In practice, the differences between them are
relatively modest, and one can generally recast a wind described
in terms of one description into the other.

%We have also included a simple kinematic description for a
%spherical wind, specifically that of \cite{castor1979}, in which
%\begin{equation}
%v=v_{o}+(v_{\infty}-v_{o})
%\left(1-\frac{R_{wd}}{r}\right)^{\beta}. \label{star_vel}
%\end{equation}
%Here, the mass-loss rate $\dot{M}_{wind}$, the terminal velocity
%$v_{\infty}$, and the acceleration exponent $\beta$ as the
%relevant wind parameters.  In principle, this allows contact with
%the large body of literature that exists for spherical winds as
%well to allow comparisons with the spectra of stellar winds. In
%practice, however, the fact that we make use of a cylindrical
%coordinate system limits the fidelity of any such comparison since
%the spatial resolution of our grid within a few $R_{wd}$ of the
%surface of the star is rather coarse.

\subsection{Radiation Sources\label{rad_source}}

The accretion disk, the WD, a boundary layer and the wind are all
sources of radiation in the code.   The disk and the WD are
treated as opaque.  Irradiation of the disk and WD by each other
and/or by the wind (``backwarming'') is not included in the
current version of the code. The wind itself emits both radiation
due resonantly scattered photons and radiation due to the fact
that the wind has been heated by other sources of emission.

The companion star is not included as a radiation source.
Including the companion star as a radiation source would require
one to adopt a full 3-d grid for the wind, since in that case the
radiation field and consequently the physical conditions in the
wind would vary with azimuth. The companion is included as a dark
absorber (located outside the wind) so that the code can be used
to model line profiles through an eclipse.

The disk, with an inner radius of $R_{wd}$ and outer radius $R_D$,
is assumed to be geometrically thin and optically thick and to
follow the standard radial effective temperature distribution
\cite[see, e.g.][]{wade1984}.

\begin{equation}
        T_{eff}(R) \; = \; T_{*} \; { \left( \frac {R_{wd}} {R} \right)}^
                {3/4} \; {\left( 1 - \sqrt{ \frac {R_{wd}}{R} }
                \right)}^{1/4},
\label{eq-disk}
\end{equation}
where $R_{wd}$ is the WD radius. $T_{*}$ is a function of the mass
accretion rate, $\dot{M}_{acc}$, and the WD mass and radius, or
more specifically,
\begin{equation}
T_* = \left(\frac{3GM_{wd} \dot{M}_{acc}}{8 \sigma \pi
R_{wd}^{3}}\right)^{1/4}, \label{eq:tstar}
\end{equation}
with $\sigma$ being the Stefan-Boltzmann constant. The disk emits
as either an ensemble of blackbodies (BBs) or an ensemble of
\cite{kurucz1993} and/or \cite{hubeny1988} model stellar spectra
\citep[SAs; see, e.g.\ ][]{wade1984,long1994}.  When assumed to
emit as a stellar atmosphere, the prescription for the effective
gravity suggested by \cite{herter1979} is utilized:
\begin{equation}
g(r)= g_o \left(\frac{r}{R_{wd}}\right)^{-15/8}
\left[1-\left(\frac{r}{R_{wd}}\right)^{-1/2}\right]^{1/8}
\end{equation}
where
\begin{equation}
g_o = 5.96\times10^5 \left(\frac{M_{wd}}{\MSOL}\right)^{5/8}
\left(\frac{R_{wd}}{10^{9}\:cm}\right)^{-15/8}
\left(\frac{\dot{M}}{10^{16}g\:s^{-1}}\right)^{1/8} \:cm\:s^{-1}.
\end{equation}
The broadening effect of the disk's Keplerian rotation on the
emitted spectrum is properly included.

The WD is taken to be spherical and to have a temperature of
$T_{wd}$. Its total luminosity is $L_{wd} = 4 \pi \sigma R_{wd}^2
T_{wd}^4$.  Its spectrum can be generated as a blackbody, or, more
commonly, from a stellar atmosphere.

Radiation due to the boundary layer (BL) between the disk and the
WD is included in a simple fashion, and is specified in terms of a
BL temperature, $T_{BL}$, and luminosity, $L_{BL}$. The size and
extent of the boundary layer is not well known on observational or
theoretical grounds. As a first approximation, we have assumed the
BL to radiate from the entire surface of the WD. Usually, the BL
is modeled as a BB, although it, too, can be described by a
stellar model atmosphere, provided its temperature is low enough.

The wind both absorbs and radiates in our code, as will be
discussed more fully in Section \ref{rad_transfer}. Each wind cell
emits free-free, free-bound and bound-bound radiation, in
accordance with its current ionization state and temperature.

All photons (energy packets) in the code are assigned a direction
($\theta,\phi$), weight ($w_{p}$) and frequency ($\nu$) at the
time of emission. Since the disk, WD, and BL all are assumed to be
optically thick, they emit photons into the ``outward'' hemisphere
only. Correspondingly, the probability, $p(\theta,\phi)$, that one
of these photons is emitted in a direction ($\theta,\phi)$) {\em
relative to the local outward normal} is proportional to
\begin{equation}
p(\theta,\phi) d\theta d\phi = \eta(\theta)
\sin{\theta}\cos{\theta} d\theta \times d\phi
\end{equation}
where $\eta(\theta)$ is the limb darkening.
% As far as I can tell we have a=0.5 and b=1.5.  This
% differs I believe from what is in Motz
For ``solid'' surfaces, we assume linear limb darkening, viz.
\begin{equation}
\eta(\theta)= a(1+b \cos{\theta}).
\end{equation}
Here, the parameter $b$ defines the strength of the
limb-darkening, which we take in the usual Eddington approximation
to be 3/2. The parameter $a$ is a normalization constant that can
be included either explicitly or implicitly. (The only important
point in the MC approach is that the total probability must sum to
unity.) The Eddington approximation is known to provide a fairly
good description of limb darkening on the sun \cite[see,
e.g.][]{mihalas1978}. The same approximation was used by SV93 and
KWD95 in their CV wind codes, and by us \cite[see,
e.g.][]{knigge1997} in fitting the disk continua of high state
CVs.

%Photons, emitted or scattered, by the wind are emitted
%isotropically, allowing for the doppler shifts that arise from the
%fact that wind is in motion. We have experimented with
%descriptions that relax this criterion, e.g descriptions in which
%scattered photons scatter preferentially along the direction of
%greatest velocity gradient, but the differences between this and
%isotropic scattering is small.  (Whether this insensitivity would
%obtain in a non-Sobolev calculation is however quite debatable.)

Each photon bundle is initially given a weight ($w_p$) that
corresponds to a fraction of the physical luminosity, or, more
usually, the band-limited luminosity (within wavelength or
frequency limits) of the system. As a result, both the
normalization and spectral shape of models generated with the code
have meaning. In the simplest case when all photon bundles have
the same weight, $w_p = L'_{tot}/N_{tot}$, where $L'_{tot}$ and
$N_{tot}$ are the sum of luminosities and number of photons
emitted by the disk, WD, boundary layer and wind in one complete
iteration. The number of disk, WD, BL, and wind photons emitted in
one complete iteration is given by $N_{disk}=N_{tot} L_{disk} /
L'_{tot}$, $N_{wd}=N_{tot} L_{wd} / L'_{tot}$, $N_{BL}=N_{tot}
L_{BL} / L'_{tot}$ and $N_{wind}=N_{tot} L_{wind} / L'_{tot}$,
respectively. Note that $L'_{tot}$ as defined is not the
luminosity that would be measured by external observers, but
rather the sum of the luminosities of the individual radiation
sources. Some of the the energy contained in these photons will be
absorbed by the wind. Indeed, the wind must (and does ignoring a
small amount of adiabatic cooling) absorb as much energy as it
emits.   As one would hope, the actual luminosity of the system as
calculated by summing the weights of all photons that emerge from
the system is quite close to the sum of the $L_{disk}$, $L_{wd}$,
and $L_{BL}$.  The only energy that is not properly accounted for
currently in the code is due to the small fraction of photons that
are ``lost'' due to impact on one of the ``solid'' surfaces before
emerging from the system.

In order to ensure that the temperature and ionization structure
of the wind converges, a statistically significant number of
ionizing photons must pass through each grid cell during the
ionization/thermal balance stage of the MC calculation (see
section \ref{balance} below). In the absence of a luminous
boundary layer, the fraction of the luminosity produced in photons
with energies in energies above the He II edge at 54.4 eV is
small, since the effective temperature of a typical CV disk is of
order 50,000 K. For this reason, we split frequency space into
multiple bands, and require that the numbers of photon bundles
generated in each band exceed certain minimum fractions. For the
calculations discussed here for example we have four bands, with
breaks at 13.6, 24.6 and 54.4 eV corresponding to the ionization
potentials of H I, He I and He II, and we require that at least
10\% of the photon bundles are in the three higher energy bands.
This requires a straightforward adjustment of the photon weights
in each band, with the largest change in the high energy band
where the number of photons was increased over that which would be
produced via uniform sampling.

Cumulative distribution functions (CDFs), defined by
\begin{equation}
P(x)=\int_{x_{min}}^{x} p(x') dx',
\end{equation}
where $p(x')$ is the probability density for a process, are used
extensively in generating and following photons within the code.
CDFs allow one to decide the source of the photon, the location of
the photon on that source, the direction of the photon, and the
frequency of the photon, where the photon is scattered, etc. For
example, for generating the frequency of a photon packet from a
specific SED, the CDF is simply
\begin{equation}
P(\nu)=\frac{\int_{\nu_{min}}^{\nu} L_{\nu} (\nu')
d\nu'}{\int_{\nu_{min}}^{\nu_{max}} L_{\nu} (\nu') d\nu'}.
\label{cdf_sed}
\end{equation}
By definition, all CDFs increase monotonically from zero to one
and can thus be easily inverted to yield appropriately distributed
random frequencies. In our code, we create an array of dimension
$n$ that contains the frequencies at which $P(\nu)$ reaches values
of $1/n$. We then generate a random number $p$ from a uniform
distribution from 0 to 1, locate the array elements that bracket
$p$, and then linearly interpolate between the associated
frequencies to get the ``exact'' frequency of the photon to be
generated.  It is also straightforward to extend this approach to
assure that certain specific frequencies, e.g. photoionization
edges, are included in such arrays. The  exact way in which CDFs
are implemented depends on the process in question. In some cases,
e.g. for determining the optical depth of scattering, the integral
can be calculated in closed form.  In others, such as in
generating photons from a BB, the CDF is constructed numerically
(in units of $h\nu/kT$) and used throughout the subsequent
calculation.

Wind photons are the most computationally intensive, because a
small number of photons is generated in each cell and because the
emission processes, especially line emission, are particularly
complex. Our approach is to store the bound-free, free-free, and
bound-bound luminosities within the appropriate frequency limits
of each cell each time we calculate a new ionization structure for
the wind. In addition, we store a very coarse CDF for the
bound-bound emission, namely the frequencies at which the CDF
reaches 0.1, 0.2, etc. We also store the band-limited luminosities
for each separate recombination process. When a wind photon must
be generated, based on the fraction of the $L_{tot}$ comprised by
the wind, we first select the cell and type -- bound-bound,
bound-free, free-free -- of photon based on the corresponding
band-limited luminosities of each process in the individual cell.
If  a free-free or free-bound photon must be generated, we use the
process described by equation \ref{cdf_sed}.  For line photons,
where linear interpolation is clearly inappropriate, the analogous
sum over a frequency ordered list of lines is
\begin{equation}
P(\nu_i)=\frac{\sum_{i'=o,i} L_{\nu_{i'}}}{L_{lines}},
\label{cdf_line}
\end{equation}
where here $L_{\nu_{i'}}$ refers to the luminosity of a specific
line within a specific cell in the grid, and $L_{lines}$ is the
band-limited luminosity due to lines in that same cell.

% The paragraph above is more accurate than the one below -- ksl
%The frequency with which
%a given primary or secondary photon is emitted depends on the spectral
%shape of the relevant radiation source. In  the MC approach, this is
%handled by constructing
%cumulative distribution functions (CDFs) for every desired SED. These
%CDFs are constructed numerically and give the probability that a
%photon is emitted with frequency $\nu < \nu\prime$ as a function of
%$\nu\prime$. By definition, all CDFs increase monotonically from
%zero to one and can thus be easily inverted to yield appropriately
%distributed random frequencies. More specifically, the frequency of
%a given photon is found by drawing a random number in the range zero
%to one from a uniform distribution and then using the relevant CDF as
%a look-up table.

%\subsection{Ionization Balance of Metals}

%The following heavy elements are included in the MC calculations: C,
%N, 0, Ne, Na, Mg, Al, Si, S, Ar, Ca, Fe, Ni. Their ionization balance
%is calculated more approximately than that of H and He by using the
%formalism developed by REF.

\subsection{Ionization and Thermal Balance\label{balance}}

Given the appropriate atomic data, an arbitrary set of elements
can be included in our MC calculations (see Section
\ref{atomicdata}). Ionization balance in the code is carried out
approximately, usually using the formalism developed by
\cite{mazzali1993}. This method is basically an improvement on the
standard nebular approximation \citep[e.g.][Eqn
5-46]{mihalas1978}.

Specifically, \cite{mazzali1993} suggest the ionization formula
\begin{equation}
\frac{n_{j+1} n_e}{n_j} = W [\xi + W(1-\xi)]
\left(\frac{T_e}{T_R}\right)^{1/2}
\left(\frac{n_{j+1}n_e}{n_j}\right)^*_{T_R}, \label{ionization}
\end{equation}
which, in principle, accounts for ionizations from and
recombinations to all levels of an ion. In this equation, the last
(``starred'') term on the right hand side refers to the Saha
abundances, $W$ is an effective dilution factor, $\xi$ is the
fraction of recombinations going directly to the ground state, and
$T_R$ and $T_e$ are the radiation and electron temperature
respectively. \footnote{For diagnostic purposes, the code also
implements several other approaches to ionization structure of the
wind, including an option to fix the abundances of individual ions
and an option to force LTE ionization populations.}

Following \cite{mazzali1993}, we define the radiation temperature
$T_R$ by computing the specific intensity-weighted
($J_\nu$-weighted) mean photon frequency, $\overline{\nu}$, for
each cell in each iteration. For a black-body radiation field
described by temperature $T_R$, this mean frequency satisfies the
relation
\begin{equation}
h \overline{\nu}=3.832~kT_R,
\end{equation}
and so
\begin{equation}
T_R = \frac{h\overline{\nu}}{3.832 k}.
\end{equation}

Both geometric dilution and absorption will cause the local
intensity, $J$, to have a value smaller than the black-body
intensity at the local radiation temperature, $B(T_R)$. Here, we
have dropped the usual subscript $\nu$ on $J$ and $B$ to denote
that we are concerned with frequency integrated quantities here.
We therefore define the effective dilution factor $W$ by demanding
that
\begin{equation}
J=W B(T_R)=W(\sigma / \pi) T_R^4.
\end{equation}
Note that in our MC approach, the frequency integrated intensity
contributed by a single photon (energy packet) to the radiation field
in a given cell, $\Delta J$, is given by
\begin{equation}
4 \pi \Delta J = \langle w_p \rangle V^{-1} l.
\end{equation}
where $\langle w_p \rangle $ is the average weight of the photon
during its passage through the cell, $l$ is the distance the
photon travels through the cell, and $V$ is the cell volume.

The quantity $\xi$ in equation~\ref{ionization} is the fraction of
recombinations going directly to the ground state. In principle,
all recombination processes should be accounted for in calculating
$\xi$. In practice, we estimate $\xi$ by using only radiative
rates calculated in the hydrogenic approximation. More
specifically, we use the analytic fits of \cite{pequignot1991},
which in turn reflect the results of \cite{martin1988}.
Physically, the first appearance of $\xi$ in the square brackets
in equation~\ref{ionization} accounts for recombinations to levels
other than the ground state, whereas the second term involving
$\xi$ accounts for photoionizations from excited levels.

There is one important difference between the calculation of the
ionization balance in our code and the prescription of
\cite{mazzali1993}. They use the ad hoc approximation $T_e = 0.9 ~
T_R$ to estimate the local electron temperature. We calculate
$T_e$ self-consistently for each cell by considering the relevant
heating and cooling processes explicitly, as discussed below. It
is worth emphasizing in this context that the thermal and
ionization equilibria are inextricably interlinked: the strength
of the heating and cooling processes (and hence $T_e$) depend on
the ionization state, but the ionization state itself also depends
on the local electron temperature. Indeed, it is this non-linear
inter-dependence that requires the use of iterative techniques in
radiative transfer codes of all kinds, including our own MC code.
The \cite{mazzali1993} approximation for $T_e$ reduces the
complexity of this inter-dependence, but (quite intentionally)
does not remove it completely: in their formalism, $T_e$ depends
on $T_R$, which depends on the ionization state, which in turn
depends on $T_e$.

The ionization and temperature structure of the wind is calculated
iteratively in an initial series of Monte Carlo calculations.  A
guess at the radiation temperature and electron temperature is
made, and this is used to calculate a trial ionization structure.
A family of typically 100,000 photons is generated from all
emission sources (disk, star, boundary layer, and wind) and
propagated through the wind. Next, the radiation temperature in
each cell is calculated, and the luminosity emitted by each cell
is compared to the luminosity absorbed by that cell. The current
estimate of the electron temperature is then adjusted, so as to
reduce the difference between emitted and absorbed luminosities.
Finally the ionization structure is updated, using this new
estimate of $T_e$. At this point, a new iteration can begin. The
whole process is repeated until both thermal and ionization
structures have converged (in the sense described below).

In the process of calculating the ionization structure, we limit
the maximum change in the electron temperature of each cell in one
iteration.  This is necessary not only because the interaction of
the heating of a cell is strongly dependent on the ionization
concentrations, which in turn reflects the electron temperature,
but also because of statistical fluctuations in the number and
frequency of photons passing through a given cell.  The number of
photons that pass through a cell depends on the effective solid
angle of that cell, which varies, and can be relatively small for
cells lie along  the cones that define the edges of the wind.
Therefore, we adaptively limit the fractional electron temperature
change in a given cycle based on the history of the cell. The
maximal temperature change allowed increases if the desired
temperature adjustment is the same direction as a previous
iteration or if the amplitude of the change is decreasing.  Cells
of this type are said to be ``converging''. Conversely, if the
desired temperature adjustment has changed sign from that of the
previous iteration and is larger in amplitude, the maximum allowed
adjustment for the cell is decreased. Limiting the rate of change
of $T_e$ help to stabilize the progression of a cell towards its
equilibrium structure. It smooths out statistical fluctuations
and prevents excessive changes in the ionization structure of
major heating or cooling constituents.

The code does not contain a formal convergence criterion; instead
we fix the number of ionization cycles at the outset and monitor
the results. In most cases, of order 20 iterations of 100,000
photons each are required to reach the point at which essentially
all of the cells are ``converging'', and in which $>90\%$ of the
cells show temperature fluctuations of less than 5\% and have
heating and cooling balanced within 10\%. At this point, the
average temperature variation of the entire wind will no longer
show a definite trend in direction in subsequent iterations, nor
will the magnitude of the adjustment exceed 100 K. If this is not
the case, we have the ability to continue the ionization
calculation from its current point (or, more commonly, to
investigate why it is not converging).

This simplified calculation of the ionization equilibrium is
clearly one of the limitations of our MC approach. We have
therefore compared the ionization equilibrium calculated by our
code to that predicted by {\sc cloudy} \citep{ferland2000} for the
simple case of an optically thin, solar abundance gas cloud with a
H density of \POW{12}{cm^{-3}} irradiated by a black body
radiation field with dilution factors W ranging from 1 to
10$^{-6}$ and radiation temperatures $T_r$ ranging from 2,000 to
1000,000 K. Fig.\ \ref{cloudy} shows one typical comparison for W
= 0.01. The overall shapes of the ionization curves, here
illustrated by He, N, C, and O are similar.
%In an attempt to
%evaluate how serious this limitation is we have as compared the
%ionization equilibrium calculated with a proper 1-dimensional
%code, Cloudy \citep{ferland2000}, to that calculated with our
%modified nebular approximation. We compared the equilibrium in the
%simple case of an single optically thin layer with a H density of
%\POW{12}{cm^{-3}} illuminated by a black body radiation field with
%dilution factors W ranging from 1 to $10^{-6}$ and temperatures
%$T_r$ ranging from 2,000 to 100,000 K. The example shown in Fig.\
%\ref{cloudy} for a dilution factor of 0.01 is typical. The overall
%shapes of the ionization curves are similar, here illustrated by
%He, N, C, and O are quite similar.
The temperatures at which an ionization state predominates are not
precisely the same, but typically the discrepancy in temperature
is less that 15\%. In this particular comparison, C IV, N V, and O
VI all peak at somewhat lower temperatures in our modified nebular
approximation than in the {\sc cloudy} calculation. On the whole,
however, given the relative crudeness of our approximation, the
degree of agreement is quite reassuring.

\subsection{Radiative transfer in the wind\label{rad_transfer}}

The fundamental equation of radiative transfer is:
\begin{equation}
\frac{dI_{\nu}}{ds}=-\kappa_{\nu} I_{\nu} +
\epsilon_{\nu}.\label{rad_trans_eqn}
\end{equation}
Here $dI_{\nu}/{ds}$ is the rate of change of the specific
intensity $I_{\nu}$ in the direction s, $\kappa_{\nu}$ is the
opacity and $\epsilon_{\nu}$ is the emissivity of the wind. In our
MC code, the second term on the right hand side of equation
\ref{rad_trans_eqn} (the wind emissivity) is accounted for by
allowing each wind cell to generate photons in accordance with the
cell's electron temperature and ionization state. The first term
on the right hand side (absorption) is implemented by
distinguishing between ``pure absorption" and ``scattering''
processes. Free-free absorption and photoionization are treated as
``pure absorption'' processes. These processes continuously remove
energy from a photon bundle as it makes its way through the wind,
but they do not change the direction of the bundle. By contrast,
electron and resonance line scattering are implemented as
scattering processes. Photon bundles scatter once their opacity
due to these processes exceeds the threshold defined in equation
1. When the bundle scatters, the direction, frequency and energy
of the bundle all change discontinuously.

%we treat free-bound and free-free processes as pure absorption
%processes, electron scattering as a pure scattering, and
%bound-bound processes as a combination of the two. This means that
%as a photon bundle traverses the wind, the weight of the photon%
%bundle is reduced due to free-free and photo-absorption, and the
%energy that is lost is deposited into the wind, but that these two
%processes do not result in any scattering. On the other, hand, the
%effects of electron scattering and bound-bound interactions on the
%photon bundle, only occur when the $\tau_{scat}$ accumulates to
%the point at which the inequality defined by equation
%\ref{scat_opt_depth} is satisfied.  It is at this point that the
%photon bundle is scattered and at this point that it can lose
%energy to the specific cell in the wind where the interaction
%occurs.

%ALTERNATIVE TO above paragraph

%In our MC code, we divide processes which affect the flow of photons
%through the wind into absorption processes and scattering
%processes. Free-free absorption and photo-ionization are treated
%as absorption processes.  As a photon moves through the wind these
%processes remove energy from the photon bundle, but do not change
%it's direction.  Thompson scattering and resonance line scattering
%are treated as scattering processes.  These processes do not
%affect the energy contained by the photon bundle until the optical
%depth due to these processes has exceeded the $\tau_{scat}$. At
%this point, the direction, frequency, and energy of the photon
%bundle are all allowed to change.

\subsubsection{Line scattering and emission}

Line transfer in our MC code is calculated in the Sobolev
approximation. In this approximation a photon travels until it
encounters a resonant surface, where the photon frequency $\nu =
\nu_{lu} (1 - v/c)$ where $\nu_{lu}$ is the frequency associated
with a particular transition and $v$ is the velocity of the wind
along the line of sight of the photon.

In the Sobolev limit, the optical depth through that surface,
allowing for stimulated emission, is given by
\begin{equation}
\delta \tau  = {(n_{l}-\frac{g_{l}}{g_{u}} n_{u})} \left(
\frac{h\nu}{4\pi} B_{lu}\right) \frac{\lambda}{|dv/ds|}
={(n_{l}-\frac{g_{l}}{g_{u}} n_{u})} \left(\frac{\pi e^2 }{mc}
f_{lu}\right) \frac{\lambda}{|dv/ds|}
\end{equation}
where $n_{l}$ and $n_{u}$ refer to the density of ions in the
lower and upper state of a particular transition, and $dv/ds$ is
the velocity gradient in the direction of photon travel
\citep[see, e.g.][]{rybicki1978}.

All levels are calculated in a two-level approximation.  The code
does not incorporate a complete model of the levels of each ion.
Specifically, for the purposes of calculating the level
populations, we use an on-the-spot approximation for each level,
i.e.
\begin{equation}
\frac{n_u}{n_l}= \frac{c_{lu}+B_{lu} W
B_\nu(T_r)}{c_{ul}+A_{ul}+B_{ul}W B_\nu(T_r)}
\end{equation}
where $A_{ul}$, $B_{ul}$, and $B_{lu}$ are the Einstein
coefficients associated with a particular transition, $c_{lu}$ and
$c_{ul}$ are collisional excitation and de-excitation rates.  In
the large J limit, the usual situation near the disk, this becomes
\begin{equation}
\frac{n_u}{n_l}= \frac{g_u}{g_l} \frac{W}{e^{h\nu/kT_r}+W-1}
\label{ratio_jlarge}
\end{equation}
The lower level population $n_1$ is calculated from the diluted
black-body approximation described by equation~\ref{lev_pop}.

When a photon bundle is scattered, it loses energy to the wind,
and is re-emitted at the frequency of the resonance transition in
the co-moving local frame. The optical depth through the resonant
surface can be very large, and therefore in a real CV wind a
photon will scatter multiple times before exiting the region where
scattering is occurring. In addition, although individual atoms
scatter isotropically (at least under the assumption of complete
redistribution), the fact that the optical depth is not the same
in all directions will cause photons to emerge from a resonant
surface anisotropically.  Here, we use an elementary escape
probability formalism to estimate both how much energy is lost to
electrons at each scattering site, and the anisotropy.
Specifically, we assume that each time a photon scatters,
collisional de-excitation reduces the weight of a photon bundle by
a fraction
\begin{equation}
q=\frac{c_{ul} (1-e^{ -h\nu/kT_e})}{A_{ul}+c_{ul} (1-e^{
-h\nu/kT_e})}\label{deexcite}
\end{equation}
where $c_{ul}$ is the collisional de-excitation rate for an ion in
the upper to lower state.  Following \cite{rybicki1983}, we assume
that the local escape probability, the probability that an emitted
photon will escape the region in a single scatter in a given
direction $\vec n$, is given by
\begin{equation}
P_{esc}=\frac{1-e^{-\tau(\vec n)}}{\tau(\vec n)} \label{pescape}
\end{equation}
where $\tau(\vec n)$ is the optical depth through the entire
resonant surface in that direction.

With this definition, it is straightforward to show that the
fraction of the energy of the photon bundle that will be absorbed
at a single scattering site is
\begin{equation}
Q=\frac{q}{q+<P_{esc}>(1-q)}
\label{deexcite_total}
\end{equation}
where $<P_{esc}>$ is the angle-averaged local escape probability.
In terms of photons weights, this means simply that following a
scatter
\begin{equation}
w_{p,out}=(1-Q) w_{p,in}.
\end{equation}

In addition to scattering line photons, the wind is allowed to
emit wind photons via thermal processes.  The effective emissivity
of the plasma is, however, reduced from that expected for an
optically thin plasma. For complete redistribution, the two-level
source function is given by
\begin{equation}
S = (1-q) <J> + q B(T_e)
\end{equation}
where q is given by equation \ref{deexcite}.  By analogy, in
extending this to multiple scattering within a single resonant
region, it is clear that  the two-level source function can be
written
\begin{equation}
S=\frac{\epsilon}{\kappa} = (1-Q) J + Q B(T_e)
\end{equation}
where Q is defined by equation \ref{deexcite_total}, and that the
effective thermal emissivity is $Q~\kappa B(T_e)$.  Thus the
effective thermal contribution due to lines is given by
\begin{equation}
4\pi \epsilon_{bb,eff} = \sum Q n_u A_{ul} h\nu
\end{equation}
where the sum is over all line transitions.  As is the case for
scattered photons, thermally emitted line photons are emitted
anisotropically in accordance with equation \ref{pescape}.

\subsubsection{Photoionization and Recombination}

As a photon bundle travels through a cell it loses energy as a
result of continuum processes.  The final weight of the photon
after accounting for continuum absorption is $w_{p,f}=w_{p,i}
e^{-\tau_{cont}}$, where $\tau_{cont}$ is the optical depth due to
all continuum processes.

The free-bound optical depth experienced by an energy packet with
frequency $\nu > \nu_{thresh}$ as it travels a distance $l$
through a grid cell is
\begin{equation}
\tau_{phot} = \kappa_{phot} l = n_{jk} \sigma(\nu)_{jk} l .
\end{equation}
where $\sigma(\nu)_{jk}$ is the photoionization cross section and
$n_{jk}$ is the number density of ions of a particular ion j in a
particular level k.  Of the energy absorbed by each bound-free
transition, a fraction $\nu_{t}/\nu$ is spent overcoming the
binding energy ($h\nu_t$) of the level of the ion, but the
remaining part, $(1-\nu_{t}/\nu)$, is taken up as kinetic energy
by the freed electron and thus goes into heating the electron
pool. Consequently, the thermal energy  gained by a grid cell (per
unit time and per transition) due to one photoionizing energy
packet passing through it is
\begin{equation}
%\G_{phot} = w_{p,i} \left(\frac{L_{tot}}{N_{tot}}\right) (1-e^{-\tau}) (1-\frac{\nu_{t}}{\nu}).
G_{phot} = w_{p,i}  (1-e^{-\tau_{cont}})
\frac{\tau_{phot}}{\tau_{cont}}(1-\frac{\nu_{t}}{\nu}).
\end{equation}

Photoionization can occur from ground or excited states.  For
photoionization calculations, $n_{jk}$ is calculated assuming
\begin{equation}
n_{jk}=n_{j} W g_k e^{-E_k/kT_r}/z_j(T_r) \label{lev_pop}
\end{equation}
where W is the usual dilution factor for that cell, $g_k$ is the
statistical weight of level k and $z_j$ is the partition function
for the ion calculated similarly. This is of course not a proper
non-LTE calculation of the level populations, but is intended as a
first-order correction to LTE.

The free-bound emissivity to the jth level of the ith ionization
state of an ion is calculated from detailed balance and given by:
\begin{equation}
4\pi \epsilon(\nu)_{jk,fb} = h\nu \left(\frac{2\pi m
k}{h^2}\right)^{-3/2} \left(\frac{8\pi}{c^2}\right)
\frac{g_{jk}}{g_e g_{j+1}} T_e^{-3/2} \nu^2 \sigma(\nu)_{jk}
e^{-h(\nu-\nu_t)/kT_e}
\end{equation}
where $g_{j+1}$ is the ground state multiplicity of the
recombining ion, and $g_{jk}$ is the multiplicity of the excited
jth level of the ion after recombination. The total rate at which
energy is lost from a grid cell due to radiative recombinations is
therefore given by
\begin{equation}
L_{fb} = V \sum_{ions} \sum_{levels} \int_{\nu_{thresh}}^{\infty}
\epsilon_{fb}(\nu) d\nu
\end{equation}
where V is the volume associated with the cell.

\subsubsection{Free-free heating and cooling}

The angle-integrated, free-free volume emissivity due to an ion j
of charge $Z$ is
\begin{equation}
4\pi \epsilon_{ff}(\nu) = n_e n_j
\frac{8}{3}\left(\frac{2\pi}{3}\right)^{1/2}\frac{Z^2 e^6}{m^2
c^3} \left(\frac{m}{kT_e}\right)^{1/2} g_{ff} e^{-h\nu /kT_e}.
\end{equation}
Consequently, the rate at which energy is lost via free-free
radiation from a grid cell in the MC simulation is just
\begin{equation}
L_{ff} = V \sum_{ions} \int_{{0}}^{\infty} \epsilon_{ff}(\nu)
d\nu.
\end{equation}
By Kirchhoff's law, the free-free absorption coefficient is
\begin{equation}
\kappa_{ff} = \frac{\epsilon_{ff}}{B_{\nu}(T_e)}.
\end{equation}
Given this, the heating rate due to free-free absorption may be
calculated analogously to that due to photoionization.
%Once again,
%a single photon with frequency $\nu$ travelling a distance $l$
%through a plasma has a probability
%\begin{equation}
%P_{ff} = 1 - e^{-\tau_{ff}}
%\end{equation}
%of being destroyed in a free-free absorption event, where
%\begin{equation}
%\tau_{ff} = \kappa_{ff} l
%\end{equation}
%is now the free-free optical depth. Consequently, an
Unlike the case of photoionization heating, all of this lost
weight represents energy that is (temporarily) being added to the
free electron pool. Thus the thermal energy (per unit time) gained
by a grid cell due to free-free absorption as an energy packet
travels through the cell is
\begin{equation}
%\G_{ff} = w_{p,i} \left(\frac{L_{tot}}{N_{tot}}\right)
G_{ff} = w_{p,i} (1-e^{-\tau_{cont}})
\frac{\tau_{ff}}{\tau_{cont}}.
\end{equation}

\subsection{Extraction of photons at specific inclination angles}

In determining the ionization structure of the wind we use a
traditional ``live-or-die'' method.  Photon bundles move through
the wind until they escape the system (``live'') or until they
encounter an opaque surface (``die'').  In spherically symmetric
situations, the live-or-die method is perfectly adequate for
constructing detailed spectra as well since all escaping photons
can be used.  If the live-or-die method is applied to our
situation however, one can use only those photon bundles that
escape in a certain narrow inclination range to construct the
spectrum for that inclination. One can use different photons to
construct spectra at a variety of inclinations in a single
calculation, but since most photons escape at intermediate
inclination angles, the photon noise in the simulated spectra will
vary significantly.

To address this problem and to allow the calculation of spectra in
eclipsing systems, we use a variation of the ``viewpoint''
technique developed by \cite{wood1991} and described by KWD95.
Specifically, we do create and follow photons throughout the grid
in a standard live-or-die fashion.  But we use these photons to
identify where in the grid photons scatter (or in some sense to
create a source function in the wind). At each place where a
photon originates or scatters, we calculate the relative
probability for a photon to be emitted or to scatter in the
specific direction of the observer. This probability is used to
increase or decrease the weight of a photon to be extracted in
that direction. For isotropic scattering in the wind, the weight
is unchanged. However, for resonance scattering, the weight is
modified by $P_{esc}\:/<P_{esc}>$, the probability that the photon
will scatter toward a specific observer divided by the
angle-averaged escape probability.  A similar correction is
applied for photons from the disk or WD photosphere, which, as
described earlier, emit anisotropically as a result of limb
darkening.

This modification of the weight accounts for local effects in
photon generation or scattering. To account for non-local effects,
the optical depth depth $\tau_{tot}$  is then calculated by
extracting the photon along the line of sight to the observer.
Assuming it does not encounter a hard surface -- the disk, the WD,
or the secondary -- the photon weight is then reduced by
$e^{-\tau_{tot}}$  and the photon is allowed to contribute to the
spectrum. In this manner, we are able to create simulated spectra
with similar amounts of photon noise regardless of the inclination
angle.  (It is similar and not identical because there are
differences in the numbers of photons that hit a solid surface.)
To verify our implementation of the viewpoint technique, we have
conducted various tests with the code that confirm that spectra
calculated using the live-or-die method and the viewpoint
technique produce the same spectral shapes.

\subsection{Atomic data \label{atomicdata}}

Our MC code incorporates a fairly flexible scheme of reading
external data files to acquire atomic data on elements, ions,
level configurations, photoionization cross-sections and
oscillator strengths.  For the sample models discussed in the
following sections, we have incorporated atomic data for H, He, C,
N, O, Ne, Na, Mg, Al, Si, S, Ar, Ca, Fe, and Ni, using the
abundances specified by \cite{verner1994}. The basic data on ions
for these elements, principally ionization potential and ground
state multiplicities were obtained from \cite{verner1996}.  For H,
He, C, N, and O, {\sc topbase} \citep{cunto1993} is the source of
both ground and excited state photoionization cross-sections; for
higher Z elements we have used analytic approximations for
ground-state populations (only) due to \cite{verner1996b}.  For
lines, we have used either the list of \cite{verner1996}, which is
a list of $\sim$5000 transitions that are ground-state connected,
or a subset of the \cite{kurucz1995} line list. In the latter
case, we abstract $\sim$55,000 lines from the entire line list
based on the oscillator strength and approximate temperature where
the associated levels are expected to be populated.  We construct
the associated levels for each ion from information contained in
the line lists, using a method similar to that described by
\cite{lucy1999b}.

\section{Comparison to KWD95 and SV93}

As indicated earlier, the two codes most widely applied to CV
spectra are those of KWD95 and SV93. Therefore it is important to
make a connection to these two codes in similar situations.

To illustrate the line shapes produced by their code, KWD95
calculated spectra for two rotating bi-conical flows with the
velocity law described by equation \ref{knigge_vel}. The two
models differed only in the degree of collimation.  Basically,
they assumed a 0.8\MSOL\ WD, accreting at a rate of \POW{-9}{\MSOL
~ yr^{1}} from a disk with a radius of 50 $R_{wd}$. They assumed
that the wind had a mass loss rate of 1\% of the accretion rate, a
velocity scale length of 10 $R_{wd}$, and a terminal velocity
1.5$\times$ that of the local escape velocity. For the weakly
collimated model, the took D to be 3.57 $R_{wd}$, which produced a
wind with opening angles ranging from 16\degr\ to 86\degr; for the
tightly collimated model they took D to be 57.12 $R_{wd}$,
implying opening angles of 1\degr\ to 41\degr.\footnote{These
correspond to models with $d/d_{min}$ of 1 and 16 in the
terminology employed by KWD95.} The complete set of parameters for
this model are summarized in Table 1. As noted earlier, the code
described by KWD95 was designed to model a few lines in detail. It
is also a Monte Carlo code, but it does not use the Sobolev
approximation and instead attempts to follow the radiative
transfer exactly.  It does not include an ionization calculation
of the wind, and so the ionization state must be specified.  For
the fiducial models described by KWD95, C IV was assumed to be the
dominant ionization state of C everywhere. In addition, thermal
emission from the wind was not included.

In an attempt to make a direct comparison between our new code and
the code developed by KWD95, we have calculated model spectra for
the parameters used by KWD95 to generate their Fig.\ 9. For this
calculation, we fixed the ionization structure and temperature of
the wind, and suppressed wind emission. We used a 100 x 100
grid\footnote{It is perhaps obvious, but each grid cell describes
cylindrical region above and below the disk, since we assume the
wind model is symmetric through the disk plane.} and a flight of
10$^6$ photons to generate CIV line profiles for the same
inclination angles shown by KWD95, i.e. 30, 60, 75 and 85\degr.

The results of this comparison are shown in Fig.\ \ref{knigge9}.
The profiles calculated with our new code are rendered in black,
while the profiles calculated with the current version of KWD95's
code are overplotted in grey. At low inclination angles, when the
observer is looking through the oncoming wind at the disk, the
profiles for the weak and strongly collimated winds exhibit strong
absorption at wavelengths shorter than the rest wavelengths of
1548.2 and 1550.8 \AA\ of the C IV doublet. There are also
emission wings near the rest wavelengths arising from scattering
in the wind. At higher inclinations the emission portion of the
profile increases in importance in both wind models.  This is due
to the decreasing projected area of the disk and exacerbated by
limb-darkening. The fact that the weakly collimated wind has
broader, but less deep, absorption profiles at low inclination and
a higher peak line-to-continuum ratio at large inclinations is due
to the fact that it is a more efficient scatterer.  It is more
efficient, because velocity gradients in the dense portion of the
wind are more varied, and the ``size'' of the wind is larger than
in the denser, but tightly collimated model.

The overall structure of the profiles is very similar, with all of
the same trends appearing in both sets of calculations. With the
exception of the highest inclination profiles (i=85\degr), both
codes produce similar line to continuum ratios, similar overall
widths, and the same narrow absorption features. At 85\degr, the
main differences are in the central portion of the profile where
KWD95 produces more flux relative to the continuum than does our
new code.  We have carried out extensive tests to verify that
there are no discrepancies in the velocity and density structures
in the two codes. We can isolate unscattered and scattered
emission in the two codes and we know that all of the profile
differences arise in the scattered light.  We suspect, but cannot
prove, that the remaining differences are due to the
simplifications inherent in the Sobolev approximation.  We know,
for example, that in our Sobolev code the scattered light profiles
are sensitive to the anisotropy of the scattering. In principle at
least, KWD95 should be expected to produce a more exact version of
the profile in multiple scattering situations.

We have also carried out a comparison to the fiducial model used
by SV93, using the parameters listed in Table 2. Briefly, this
model is for a system with a 40,000 K WD with a mass of 0.8 \MSOL\
and radius of \EXPU{6}{8}{cm}.  The disk extends from the WD
surface to 34 $R_{wd}$ and is presumed to be in a steady-state
with an $\dot{m}_{disk}$ of \POW{-8}{\MSOL \: yr^{-1}}. The wind
is constrained to emanate from the disk in a broad annulus between
4 and 12 $R_{wd}$. At the inner wind edge, streamlines are
inclined at 12\degr\ with respect to the disk normal, whereas at
the outer wind edge, the streamline inclination is 65 \degr. The
total mass loss rate in the wind is 10\% of the accretion rate.
The scale length for defining the velocity of the wind is 100
R$_{wd}$, so the wind does not reach terminal velocity until the
wind is well outside the system. The terminal velocity along each
stream line is 3 times the escape velocity at the footpoint of the
wind, so the most rapidly moving portions of the wind are located
at its inner edge.

For this particular calculation, we employed the 100 x 100 grid,
and used 20 cycles of \POW{6} photons to calculate the ionization
structure; at this point, 96\% of the cells were converged by our
criteria. Consistent with SV93, all of the sources of continuum
radiation were simulated in terms of appropriately-weighted black
bodies. We then generated \EXPN{2}{6} photons to create simulated
CIV line profiles at 10 degree intervals; these are plotted as
black lines in Fig.\ \ref{sv_c4}. For this calculation, C IV was
represented as a singlet and emission from the wind was suppressed
in an attempt to mimic SV93's calculation as closely as possible.
In this case, the line profiles at very low inclination show
emission profiles. Unlike the KWD95 prescription, the SV93
fiducial model does not have a wind covering the innermost portion
of the disk (where much of the UV emission arises) and as a result
there is little or no absorption at very low inclination.  When
the inclination reaches 30\degr, and the observer is viewing the
system from within the wind cone, absorption profiles develop. The
blue edge of the absorption is sharp, but moves redward as
inclination increases. This is due to the fact that the fastest
stream lines, which are at the inner edge of the wind, intersect
the line of sight with increasing obliquity. The emission wing
grows increasingly strong with inclination for the same reasons as
in the KWD95 model above. When the inclination reaches 70\degr,
which is beyond $\theta_{max}$, the profile is completely in
emission.

Also shown in grey in Fig.\ \ref{sv_c4} are the profiles
calculated by SV93 for this model, as digitized from their Fig.\
5. As was the case for comparisons with KWD95, the same general
features are seen in both calculations, and there is quite good
agreement in the emission line portion of the profiles; this may
reflect the fact that the SV93 escape probability assumptions are
rather similar to our own.  There is, however, a difference in the
shapes of the absorption profiles at intermediate inclination
angles (40-60\degr). The blue edges of the lines lie at about the
same wavelengths, and the edges are sharp in both sets of models.
However, the SV93 profiles exhibit maximum absorption at about
1540 \AA, with the absorption depths declining towards the blue
edge. These differences are almost surely due to the details of
the ionization structure calculated in the two cases.  The
ionization fraction of C IV that we calculate in the inner 50
$R_{wd}$ portion of the wind is shown in Fig.\ \ref{sv_c4_abun},
and should be compared with Fig.\ 4 of SV93. However, in our
calculation, the fractional abundance of CIV exceeds 0.5 over  a
significant portion of the wind at all cylindrical radii from the
WD; by contrast, in the SV93 calculation, the CIV density declines
significantly beyond 25 $R_{wd}$ in the radial direction and also
beyond the inner 5 $R_{wd}$ above the disk. As a result, the
optical depth at the blue edge of the C IV profile is much higher
in our calculation. It is not clear at this stage which ionization
structure is more correct. The SV93 ionization calculation is not
described in detail and we have adopted a rather simplified
approach to ionization ourselves. If our ionization calculation is
more accurate, it suggests that somewhat lower mass loss rates
will be required to model the line spectra of CVs.

Overall, we conclude that the agreement between our calculations
and those of SV93 is reassuring, certainly for us, and possibly
for SV93 as well.  The differences in the profile shapes we
calculate are small compared to those that will result from
current uncertainties in the basic parameters of CV winds, such as
$\dot{m}_{wind}$, the degree of collimation, and the velocity law.

\section{A typical calculation}

Our primary motivation in developing a new radiative transfer code
was to enable us to simulate spectra over a wide wavelength range.
Fig.\ \ref{sv_bb62} illustrates the results a simulation covering
1000 \AA\ in the FUV waveband for the SV93 fiducial model view at
an inclination of 62.5\degr.  For this and other examples
described below, $10^{6}$ photon bundles were created and followed
through the wind to create the final spectrum. Since both disk and
WD spectral distributions have been formed from appropriately
weighted black bodies, all of the features in the spectrum are due
to the effects of the wind, including the sharp cutoff at 912 \AA.
In this simulation, C IV and N V show  well-developed P Cygni
profiles. O VI is not evident.  A number of lower ionization
potential lines of ions, such as C III, N III, and Si IV, appear
only in absorption. These ions are present only near the disk; as
a result, we see them only in absorption along the line of sight
to the disk. Ly$\alpha$ and Ly$\beta$ are also seen in absorption,
even though the ionization fraction of neutral hydrogen is quite
small.

In Fig. \ref{sv_hub}, the same model is shown for inclinations
ranging from 10\degr\ to 80\degr, but in this case the disk and WD
are simulated in terms of appropriately weighted stellar
atmospheres. At 10\degr, when the observer views the system from
``inside'' the inner wind cone, the effects of the wind on the
spectrum are quite subtle, and most of the features are
photospheric with the exception of the narrow emission profiles of
 C IV. At angles of 27.5, 45, and 62.5\degr, the observer is
viewing the disk through larger and larger columns of wind
material, so more and more ions become apparent. Often these ions
deepen features in the photospheric spectrum, since, not
surprisingly, the ionization state of the wind immediately above
the disk is not that different from the photosphere. At 62.5\degr,
the line of sight skims over the outer lower temperature portion
of the disk, so the lower ionization lines appear in the spectrum.
The profiles of some lines (e.g. C IV and N V) already exhibit
strong emission at this inclination. This reflects that fact that
the ionization stages corresponding to these lines are highly
extended in the wind. As a result, relatively large portions of
the line forming regions are not viewed projected against the
bright disk, and these regions contribute only scattered emission.
A comparison of the spectrum in Fig.\ \ref{sv_bb62} to the
62.5\degr\ panel in Fig.\ \ref{sv_hub} shows how important an
accurate ``photospheric'' model of the disk and WD remains for
modeling the FUV spectra of CVs. Finally, at 80\degr, when the
observer views the system from ``outside'' the wind cone, the
spectrum is dominated by wind emission lines.

\section{Modeling the Spectrum of a Real
CV}

In the absence of an agreed physical model of the wind of a DN in
outburst, any description of a bi-conical flow necessarily has a
large number of parameters. It is far from clear how to explore
the parameter space, especially since the luminosity, geometry and
SED of the boundary layer is unknown in almost all CVs.

However, an example of the type of ``fit'' that can be obtained is
shown in Fig.\ \ref{zcam_knigge}, which shows data obtained with
HUT during a normal outburst of Z Cam compared to a model spectrum
generated using a wind geometry nearly identical to that described
by \cite{knigge1997}.  Qualitatively, the model spectrum and the
data resemble one another quite well. In overlaying the two
spectra, we have assumed the value of E(B-V) and NH used by
\cite{knigge1997}, and simply renormalized. The strong resonance
lines of C IV, N V, and O VI have approximately the right
profiles, and many of the features of the lower ionization lines
can be seen in both the model and in the data.  The specific model
shown here has the same amount of collimation, the same
acceleration length and terminal velocity, the same
$\dot{m}_{wind}$ of \EXPU{1}{-10}{\MSOL \: yr^{-1}}, and a similar
value of \EXPU{6}{-9}{\MSOL} for $\dot{m}_{disk}$ (well within the
range derived from fitting the continuum spectrum). The disk
continuum was generated from an ensemble of stellar atmospheres,
and the \cite{kurucz1995} line list was utilized in the wind. The
model shown is for an inclination of 69\degr, which is near the
maximum of range allowed for Z Cam. The main effect of reducing
the inclination angle slightly would be to decrease the emission
wing of C IV. The only real change from the model described by
\cite{knigge1997} is that the model shown here is has the
parameter $\alpha$ controlling the mass loss rate as a function of
radius is 1 instead of 0.4. In our model the mass loss rate per
unit area is proportional to the surface flux, which concentrates
the wind closer to the WD. This helped to reduce the strength of
several low ionization lines, which otherwise produced narrow deep
absorption lines not seen in the data.

As noted above, we did not systematically search for a best fit to
the Z Cam spectrum.  Therefore, given our limited experience at
this point, we cannot rule out that different wind structures
might produce similar emergent spectra. We did, however, generate
other models around the one described here. Small changes, in for
example the collimation of the wind or the acceleration length, do
improve individual lines, but do not qualitatively change one's
impression of the fit overall. One interesting fact is that it is
very difficult to fit both the region below 1000 \AA\ and the
resonance lines when model atmospheres are used to generate the
disk spectrum. This is because both the wind and the the turnover
in the underlying disk continuum depress the overall flux in this
region.  This is clearly an area we will have to explore in our
attempts to fit FUSE spectra in the future.

\section{Summary}

The inherent advantage of a Monte Carlo approach in simulating the
spectra of cataclysmic variables is the ease with which one can
model complex geometries.  In the Monte Carlo approach, axially
symmetric winds are not significantly more difficult to model than
spherical winds.  In the case of our code, we have now progressed
to the point of spectral verisimilitude, where it is difficult to
distinguish a simulation from data.  The wind prescriptions
required to produce verisimilitude are not that different from
those which were developed to model C IV line profiles.

On the other hand, preliminary comparisons of the models and data
suggest that considerable work will be required to reproduce the
spectra of some dwarf novae.  Nevertheless, we are already having
some successes, such as the model shown here for Z Cam and our
model of the FUSE spectrum of SS Cygni (Froning et al. 2001). We
are now embarking upon an effort to fully explore the parameter
space inherent in the existing models and to compare these models
to a variety of HUT, FUSE and HST spectra of high state CVs. This
should allow us to determine whether there is a single class of
models, broadly or narrowly collimated for example, that can be
used to approximate the wind geometry in the majority of CVs.

%It is also possible that some additional physics is crucial to the
%successful modeling of DN winds. As an example, we do not
%currently make allowances for the possibility that shocks are
%required to model the O VI line profiles that are seen in some
%CVs; shocks would not be unexpected and are commonly used to
%address O VI profiles in O star spectra.

\acknowledgments{This work was supported by NASA through grants
associated with our analysis of HST and FUSE spectra of dwarf
novae.  Specifically we acknowledge support from grants G0-7362
and GO-8279 from the Space Telescope Science Institute, which is
operated by AURA, Inc., under NASA contract NAS5-26555, as well
from grants NAG5-9283 and NAG5-10338, which are associated with
our FUSE projects.  We have also benefitted from many useful
conversations from our colleagues, most notably Janet Drew, Ivan
Hubeny, and Tom Marsh.  This work was completed during a
delightful visit by one of us (ksl) to the Department of Physics
\& Astronomy of the University of Southampton.}

%\appendix
%\pagebreak
%\include{tab1}

\pagebreak

\figcaption[fig1.sv.01mar26.ps]{A cartoon illustrating the basic
geometry of the bi-conical wind description of SV93.
\label{sv_cartoon}}

\figcaption[fig1.knigge.01mar26.ps]{A cartoon of the wind
description employed by KWD95. The wind is symmetric with respect
to the disk.\label{knigge_cartoon}}

\figcaption[cloudy_python_0.01.ps]{A comparison of ionization
fractions of He, C, Ni, and O as calculated using Cloudy and with
the modified on-the-spot approximation used within Python for a
black body radiation field with a effective dilution factor W of
0.01. Neutral species as well as He II, C IV, N V, and O VI appear
as black lines in the figure. \label{cloudy}}

\figcaption[knigge9.ps]{C IV line profiles (in black) calculated
with our code for two models using the KWD95 prescription for a
bi-conical rotating wind. In these particular calculations, the
ionization abundances were fixed, and C was assumed to be all C
IV.  The upper group of profiles is for KWD95's fiducial model of
a wind with a wide opening angle, while the lower panel indicates
the results for a highly collimated wind.  Four inclination
angles, 30 ,60, 75, and 85\degr, are shown for each model. For
comparison, the figure also shows profiles (in grey) calculated
with the ``exact'' MC code described by KWD95.  The models
selected are identical to those used as fiducial models by KWD,
and this figure is almost identical to their Fig.\ 9.
\label{knigge9}}

\figcaption[sv_c4.ps]{ C IV line profiles (in black) calculated
with our code in 10 degree increments for the fiducial model
discussed by SV93.  The profiles calculated by SV93 (as digitized
from their original figure) are plotted in grey. Since SV93 did
not include thermal emission from the wind, thermal emission was
also suppressed in our calculation; including the wind emission
does make a significant difference in the profiles. \label{sv_c4}}

%\figcaption[sv_c4_abun.eps]{Densities of Si IV (upper left), C IV
%(upper right), N V (lower left)  and O VI (lower right) in the
%standard SV93 model for the inner 50 $R_{wd}$ portion of the wind
%in our calculation of a standard SV93 model. All of the scales are
%logarithmic from \POW{4}{cm^{-3}} to \POW{8}{cm^{-3}}. Close to
%the disk C III is the dominant ion. Far from the disk (in the z
%direction) C V is the dominant ion.\label{sv_c4_abun}}

\figcaption[sv_c4_abun.eps]{Densities (left) and ionization
fraction (right) of C IV in our calculation of a standard SV93
model for the inner 50 $R_{wd}$ portion of the wind. The disk lies
along the x-axis.   For the densities, the contours correspond to
values of log(N(C IV)($cm^{-3}$)) of 2, 4, 6, 7, 8, 9. As one
might expect, the highest C IV densities occur relatively close to
disk, where the wind is moving relatively slowly.  For ionization
fraction, the contours represent concentrations of 10, 20, 30, 40,
50, 60, 70, 80 and 90\%.  The ionization concentrations is high
over a relatively extended region peaking about 30 WD radii from
the center of the system. On the edge of the wind closest to the z
axis, the dominant ionization stage is C V, while close to the
disk plane C III tends to predominate. \label{sv_c4_abun}}

%\figcaption[sv_spec_tot.ps]{The emergent angle-averaged spectrum
%during the last ionization cycle. This is not in the text at
%present}

\figcaption[bb_bw.01sept26.eps]{The calculated spectrum for the
fiducial SV93 geometry as observed at an inclination of 62.5
\degr\ and assuming both the disk and the WD radiate as
blackbodies.\label{sv_bb62}}

%\figcaption[fig.scat.01mar27.ps]{A representation of the sites at
%which photons scatter in a standard SV93 model. This is not in the
%text at present \label{scat_sites}}

\figcaption[sv_hub.ps]{The calculated spectrum for the fiducial
SV93 geometry, which has $\dot{m}_{disk}$ of \POW{-8}{\MSOL \:
yr^{-1}} and $\dot{m}_{wind}$ of \POW{-9}{\MSOL \: yr^{-1}} as a
function of inclination angle. In contrast to Fig.\ \ref{sv_bb62},
here the disk and WD were simulated in terms of appropriately
weighted Hubeny (1988) model atmospheres.\label{sv_hub}}

\figcaption[zcam57_bw.01sept26a.eps]{A comparison between the HUT
spectrum of Z Cam (in black) and a model spectrum (in grey)
calculated using a prescription for the wind geometry similar to
that suggested by Knigge et al.\ (1997). The model has been
smoothed slightly to reflect the resolution of HUT.  Some of the
stronger lines, including the Lyman series of H are labelled. The
regions near strong features at 1304 \AA\ and 1356 \AA\ may have
been affected by air glow lines.\label{zcam_knigge} }

\clearpage
\begin{center}
\begin{deluxetable}{lcc}
\tablecaption{Fiducial KWD Model of Bi-conical Wind }
\tablehead{\colhead{Parameter} & 
 \colhead{Value} & 
 \colhead{Units} 
}
\scriptsize
\tablewidth{0pt}\startdata
$M_{wd}$ &  0.8 &  \MSOL \\ 
$R_{wd}$ &  \EXPN{6.8}{8} &  {cm} \\ 
$T_{wd}$ &  0 &  K \\ 
$T_{bl}$ &  0 &  K \\ 
$L_{bl}$ &  0 &  $ergs~s^{-1}$ \\ 
$\dot{M}_{disk}$ &  \EXPN{1}{-9} &  $\MSOL~yr^{-1}$ \\ 
$R_{disk}$ &  50 &  $R_{wd}$ \\ 
$\dot{M}_{wind}$ &  \EXPN{1}{-11} &  $\MSOL~yr^{-1}$ \\ 
D &  14.3,~57.1 &  $R_{wd}$ \\ 
$D/D_{min}$ &  1,~16 &  - \\ 
$R_{v}$ &  10 &  $R_{wd}$ \\ 
$v_\infty$ &  1.5 &  $v_{esc}$ \\ 
$\alpha$ &  1 &  - \\ 
$\beta$ &  0.8 &  - \\ 
\enddata 
\end{deluxetable}
\end{center}

\newpage
\begin{center}
\begin{deluxetable}{lcc}
\tablecaption{Fiducial SV Model }
\tablehead{\colhead{Parameter} & 
 \colhead{Value} & 
 \colhead{Units} 
}
\scriptsize
\tablewidth{0pt}\startdata
$M_{wd}$ &  0.8 &  \MSOL \\ 
$R_{wd}$ &  \EXPN{7}{8} &  {cm} \\ 
$T_{wd}$ &  40,000 &  K \\ 
$T_{bl}$ &  10,000 &  K \\ 
$L_{bl}$ &  0 &  $ergs~s^{-1}$ \\ 
$\dot{M}_{disk}$ &  \EXPN{1}{-8} &  $\MSOL~yr^{-1}$ \\ 
$R_{disk}$ &  34 &  $R_{wd}$ \\ 
$\dot{M}_{wind}$ &  \EXPN{1}{-9} &  $\MSOL~yr^{-1}$ \\ 
$\theta_{min}$ &  20 &  \degr \\ 
$\theta_{max}$ &  65 &  \degr \\ 
$r_{min}$ &  4 &  $R_{wd}$ \\ 
$r_{max}$ &  12 &  $R_{wd}$ \\ 
$R_{v}$ &  100 &  $R_{wd}$ \\ 
$v_\infty$ &  3 &  $v_{esc}$ \\ 
$\alpha$ &  1.5 &  - \\ 
$\lambda$ &  0 &  - \\ 
\enddata 
\end{deluxetable}
\end{center}

\newpage
\pagestyle{empty}
\begin{figure}
%\plotone{sv.cartoon.ps}
\plotone{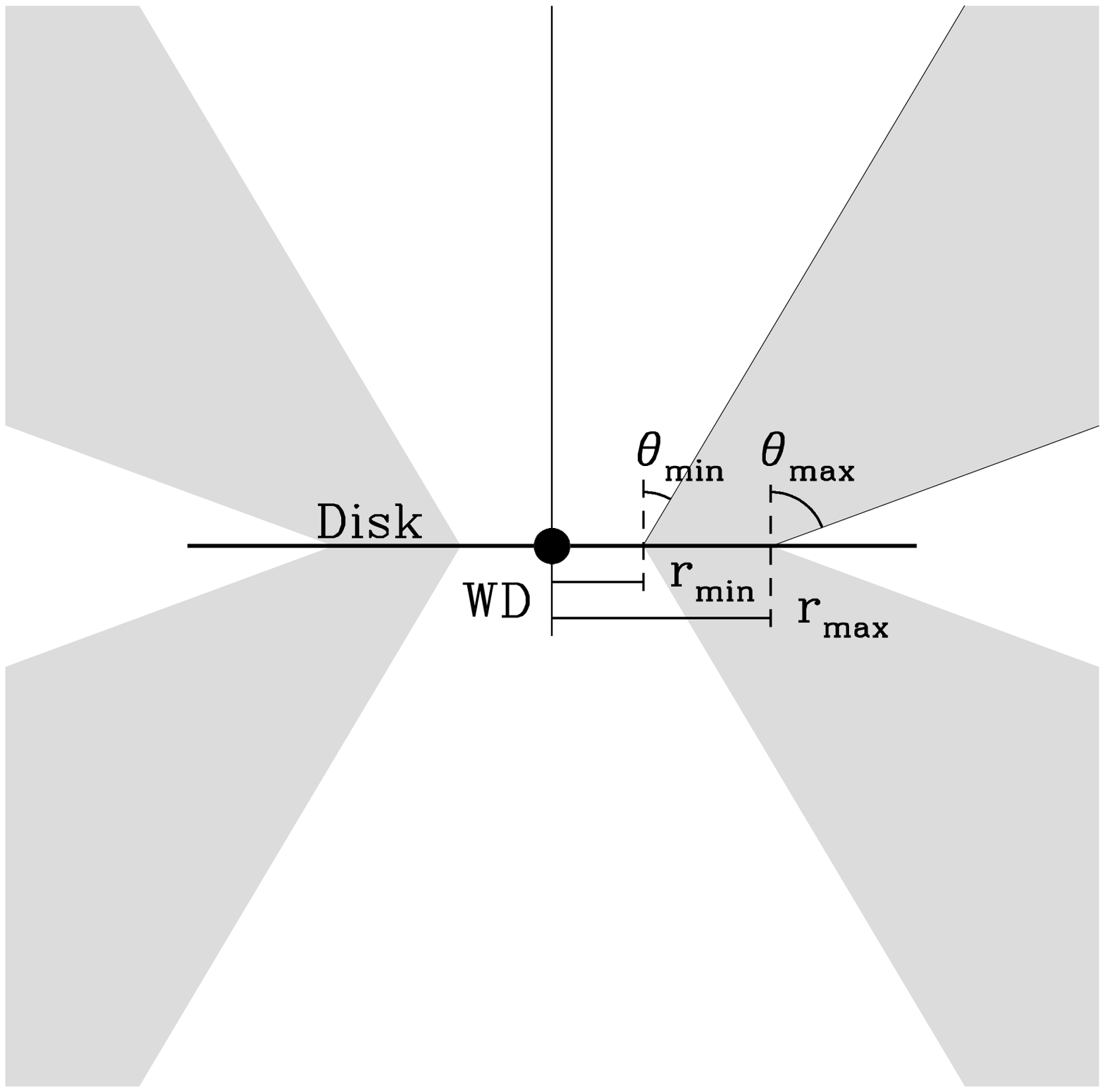}

\end{figure}

\newpage
\pagestyle{empty}
\begin{figure}
%\plotone{knigge.cartoon.ps}
\plotone{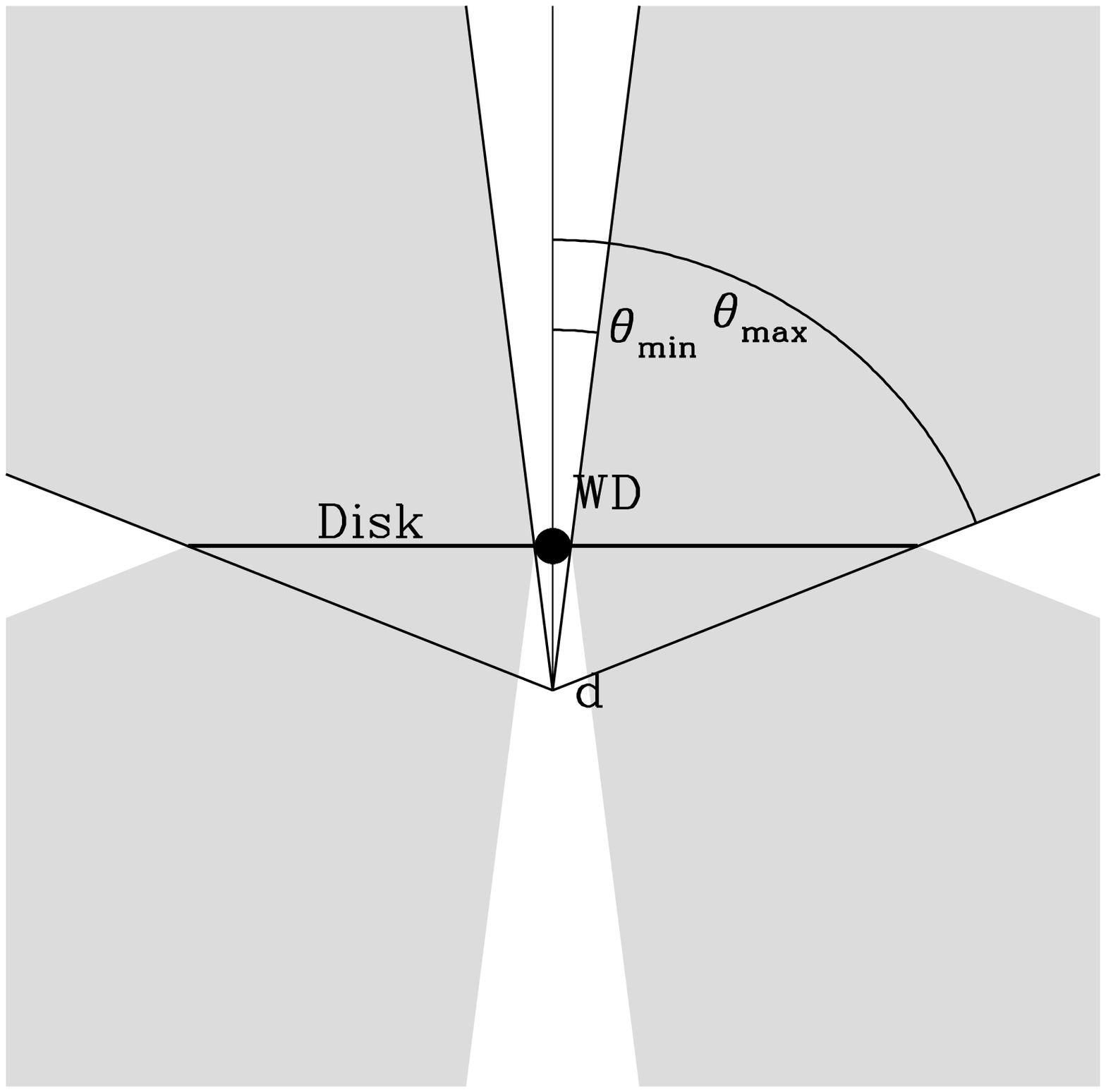}
\end{figure}

\begin{figure}
%\plotone{cloudy_python_0.01.ps}
\plotone{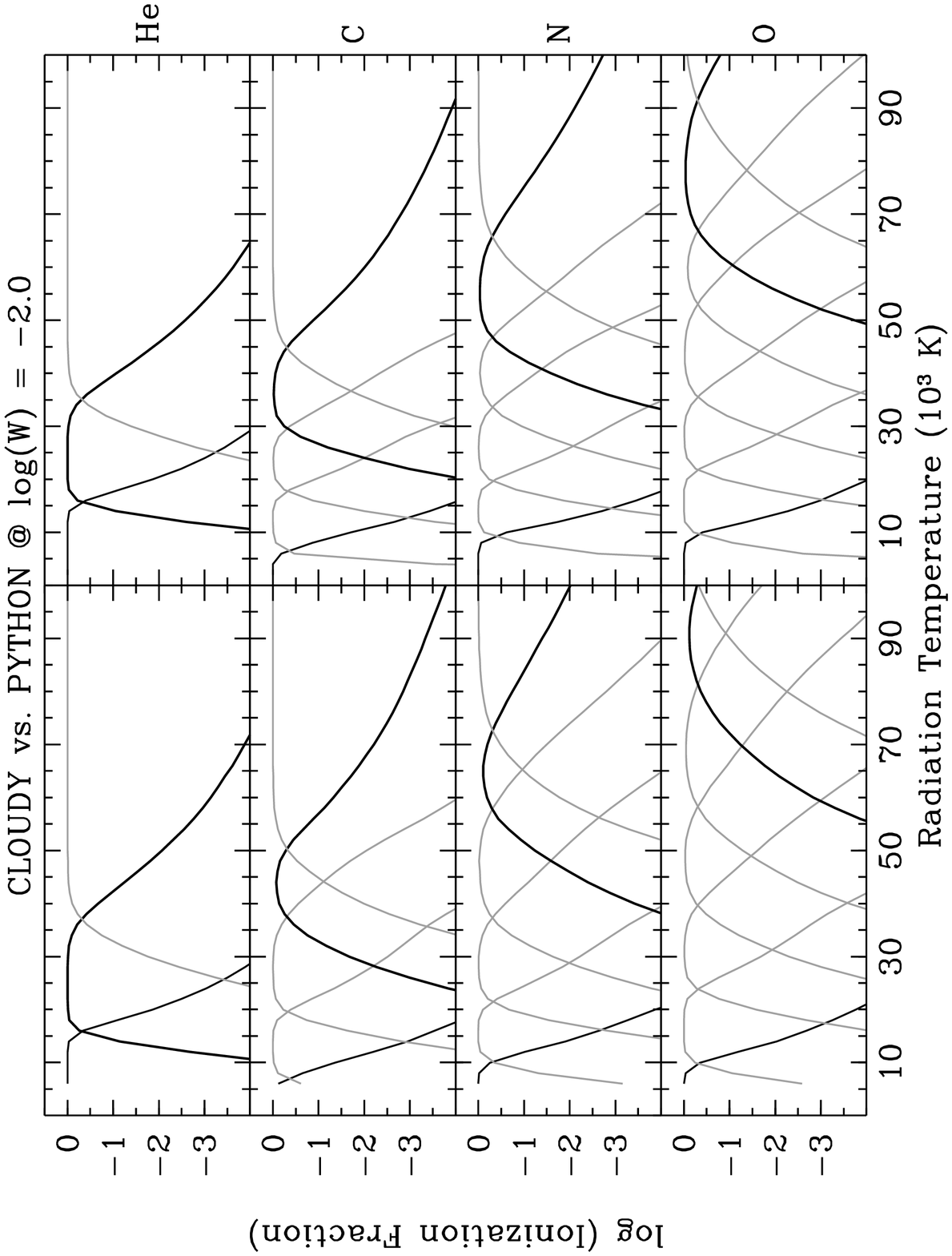}
\end{figure}

\begin{figure}
%\plotone{knigge9.ps}
\plotone{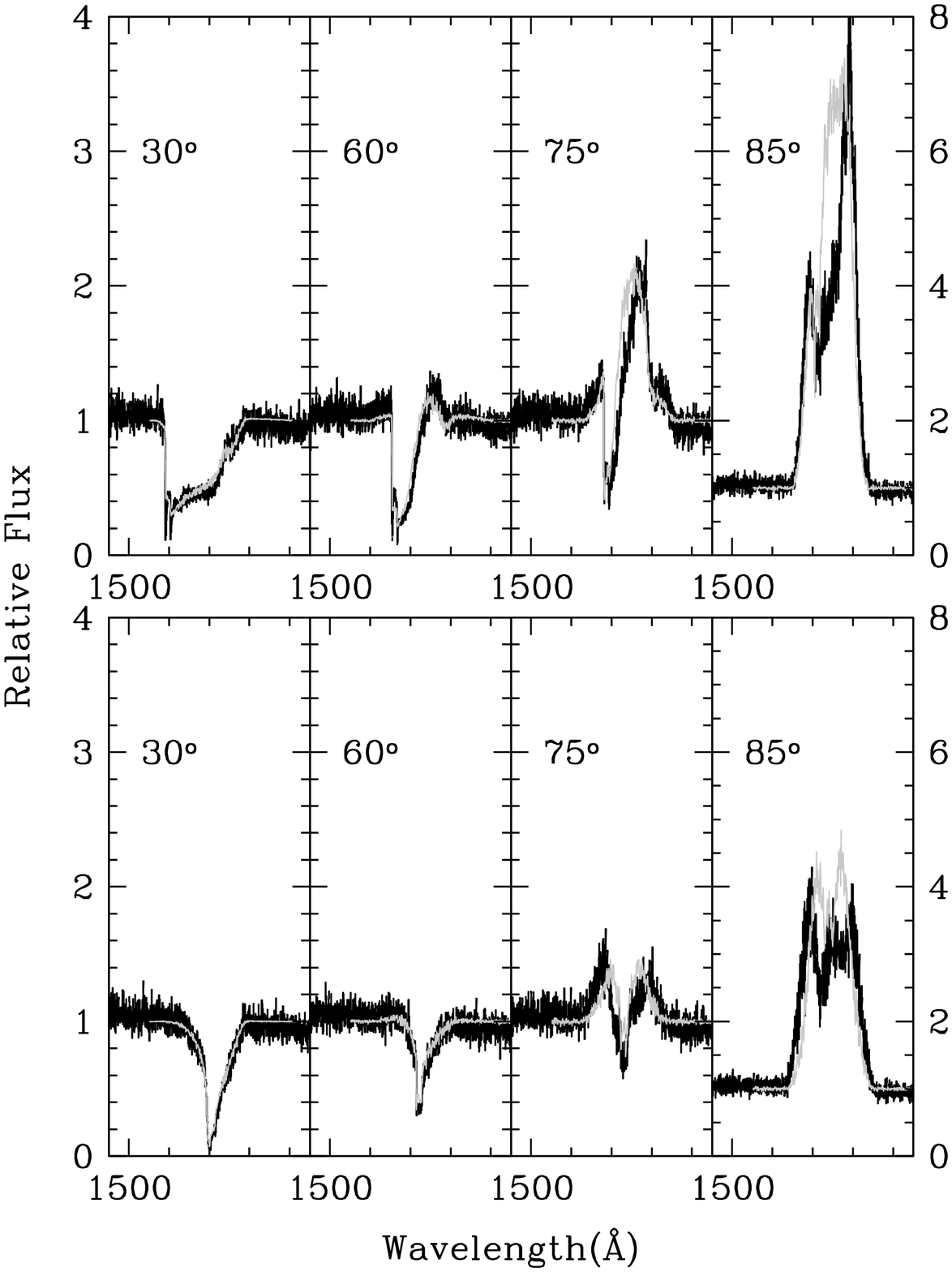}
\end{figure}

\begin{figure}
%\plotone{sv_c4.ps}
\plotone{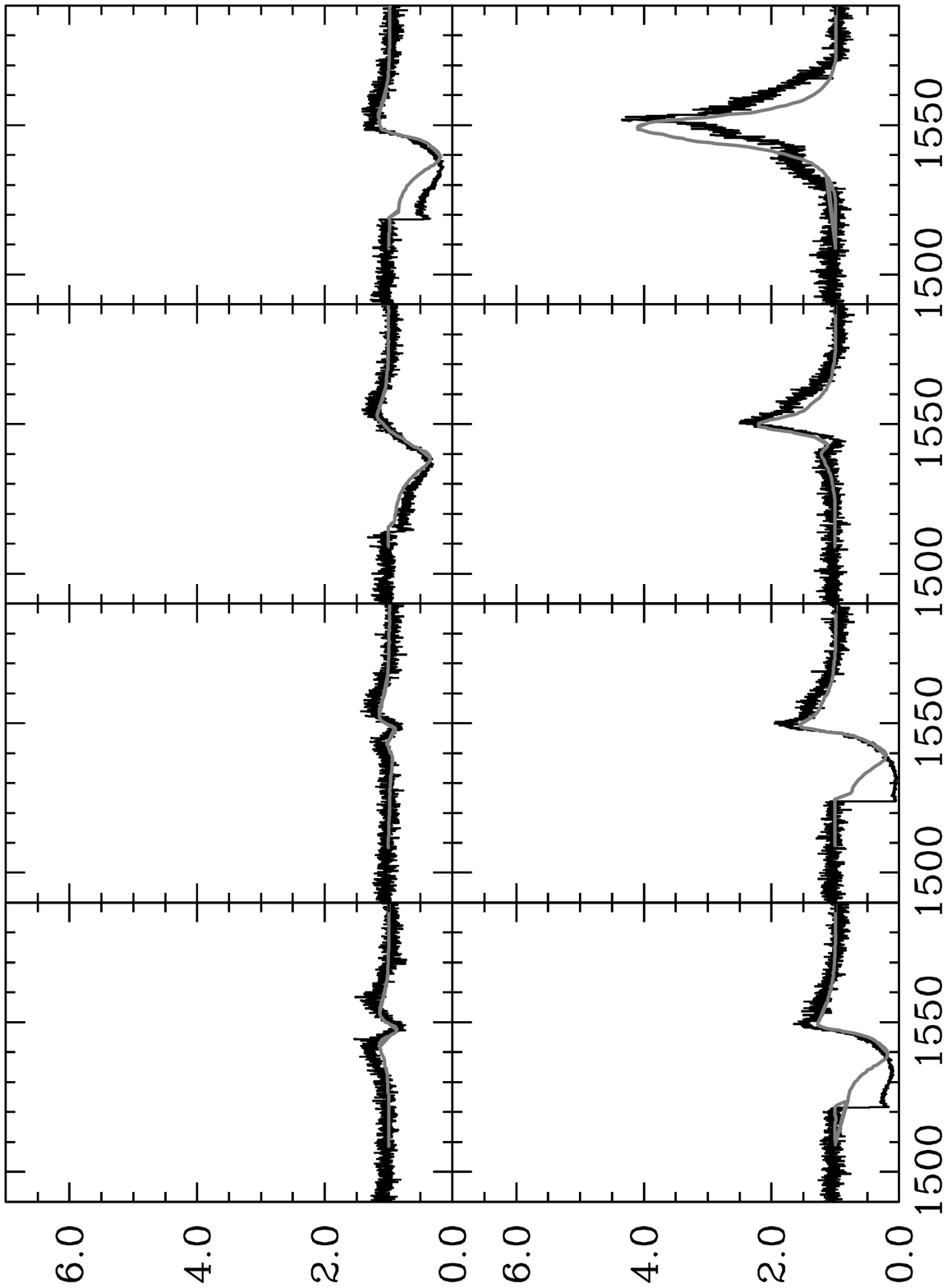}
\end{figure}

\begin{figure}
%\plottwo{c4_ioniz_con_fin.ps}{c4_ioniz_fin.ps}
\plottwo{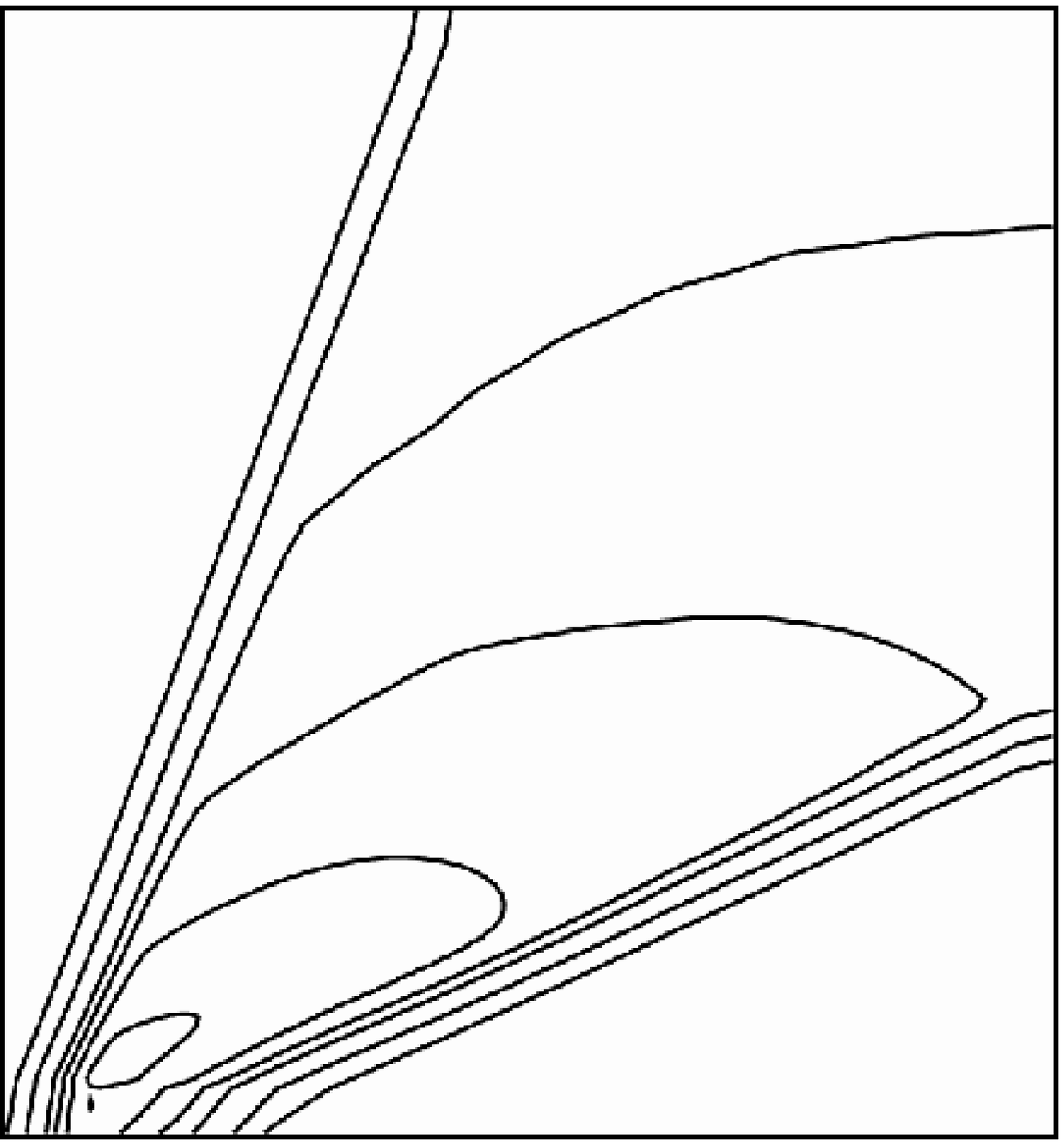}{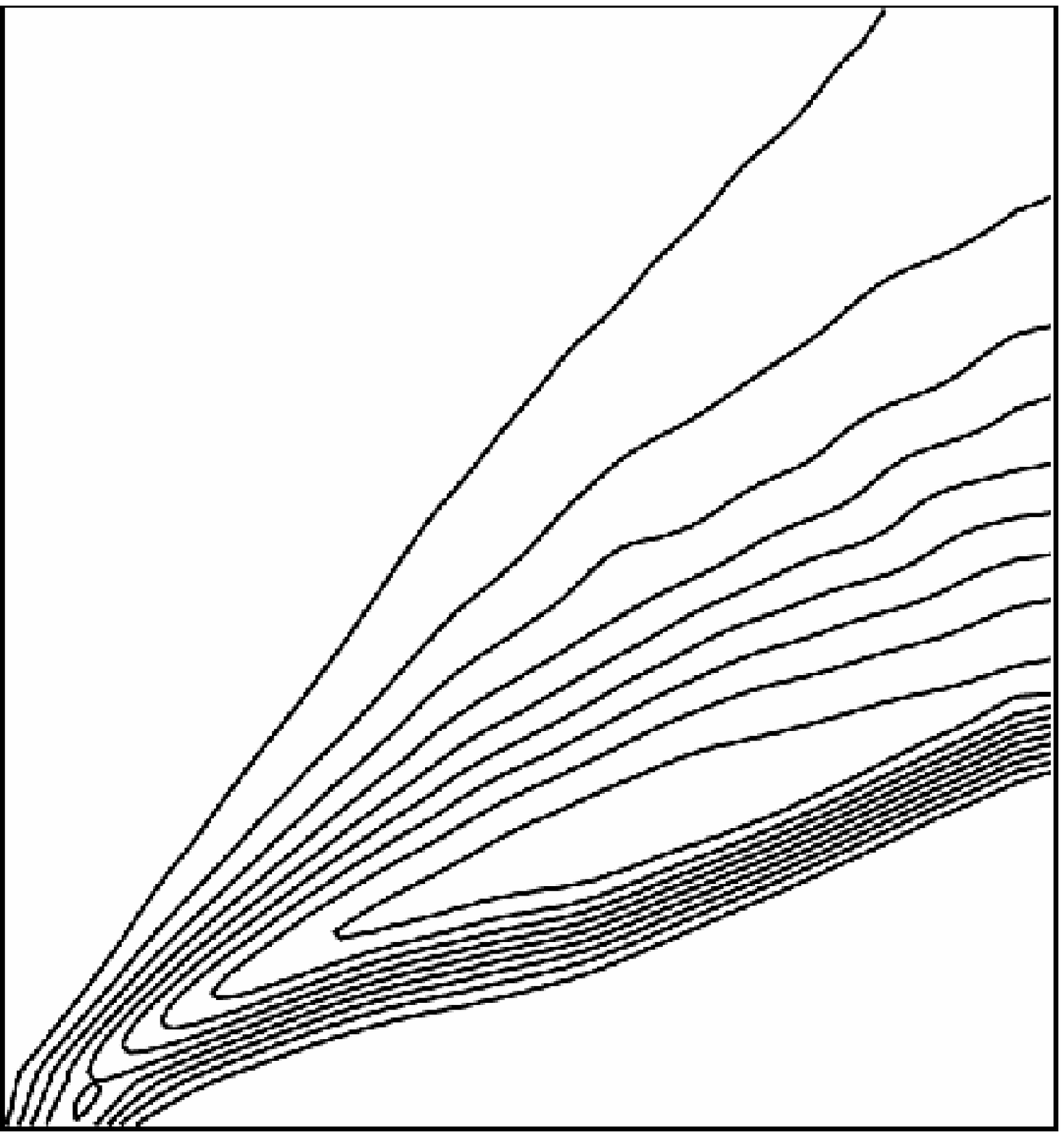}
\end{figure}

\begin{figure}
%\plotone{sv_bb62.ps}
\plotone{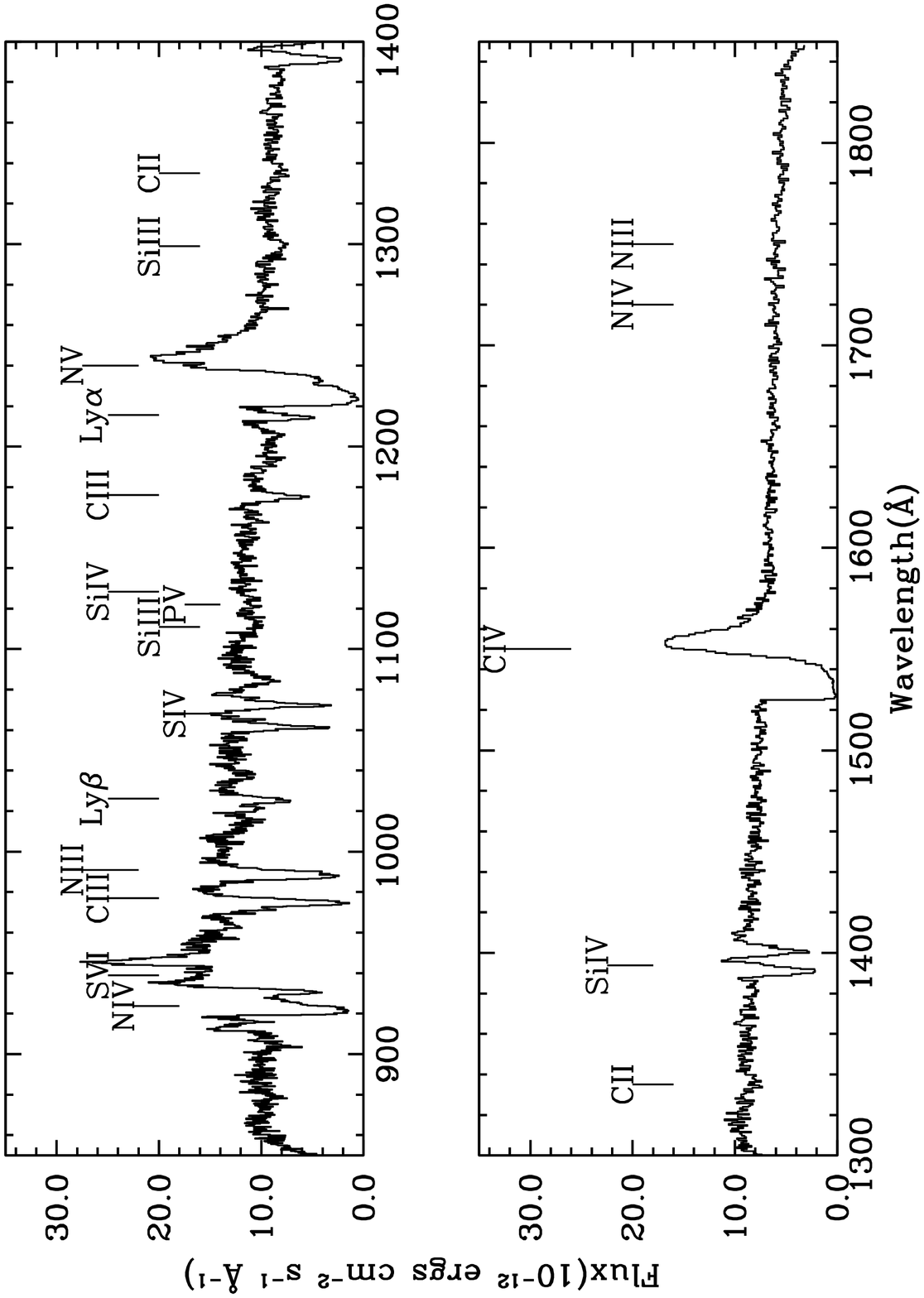}
\end{figure}

\begin{figure}
%\plotone{sv_hub.ps}
\plotone{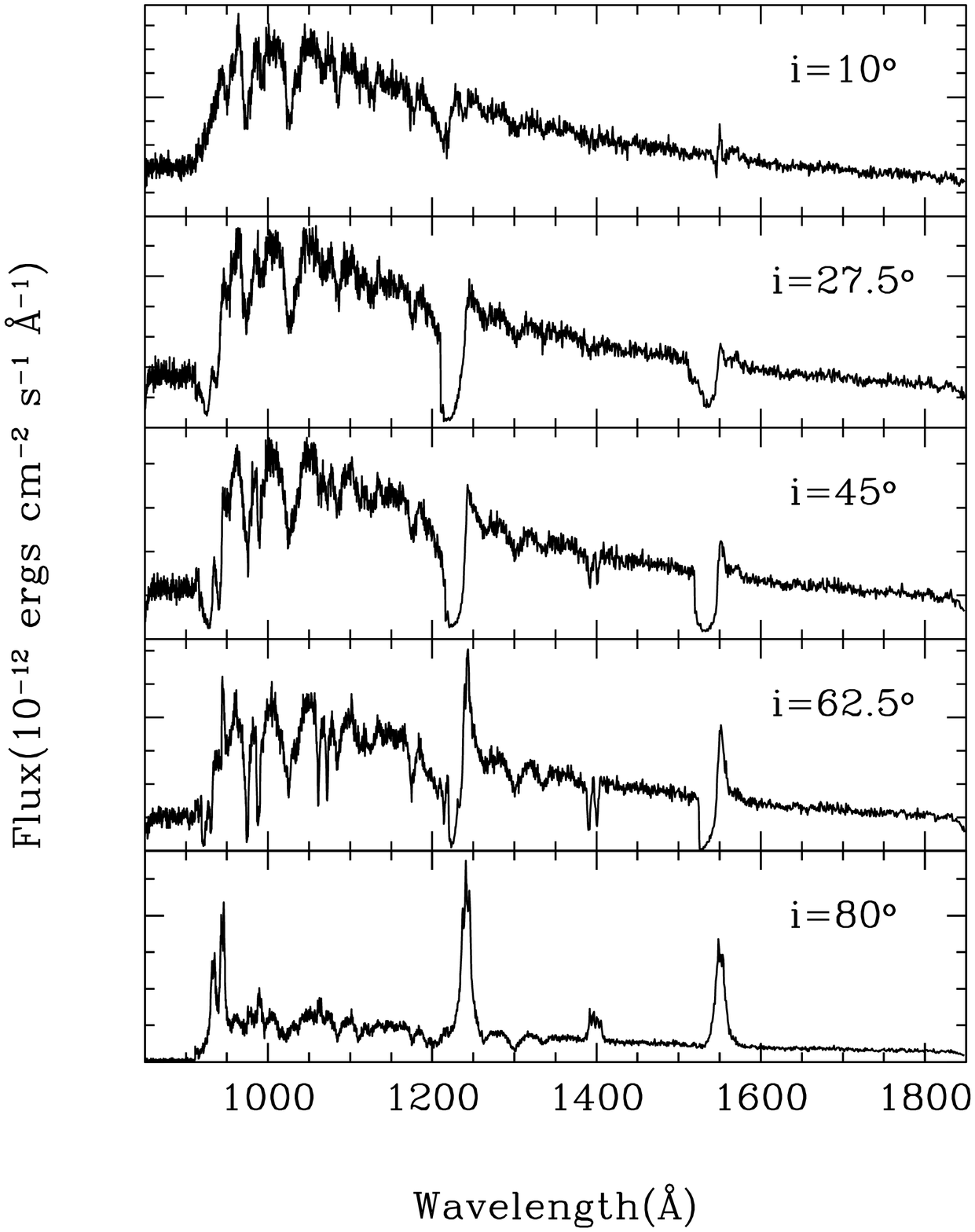}
\end{figure}

\begin{figure}
%\plotone{zcam.ps}
\plotone{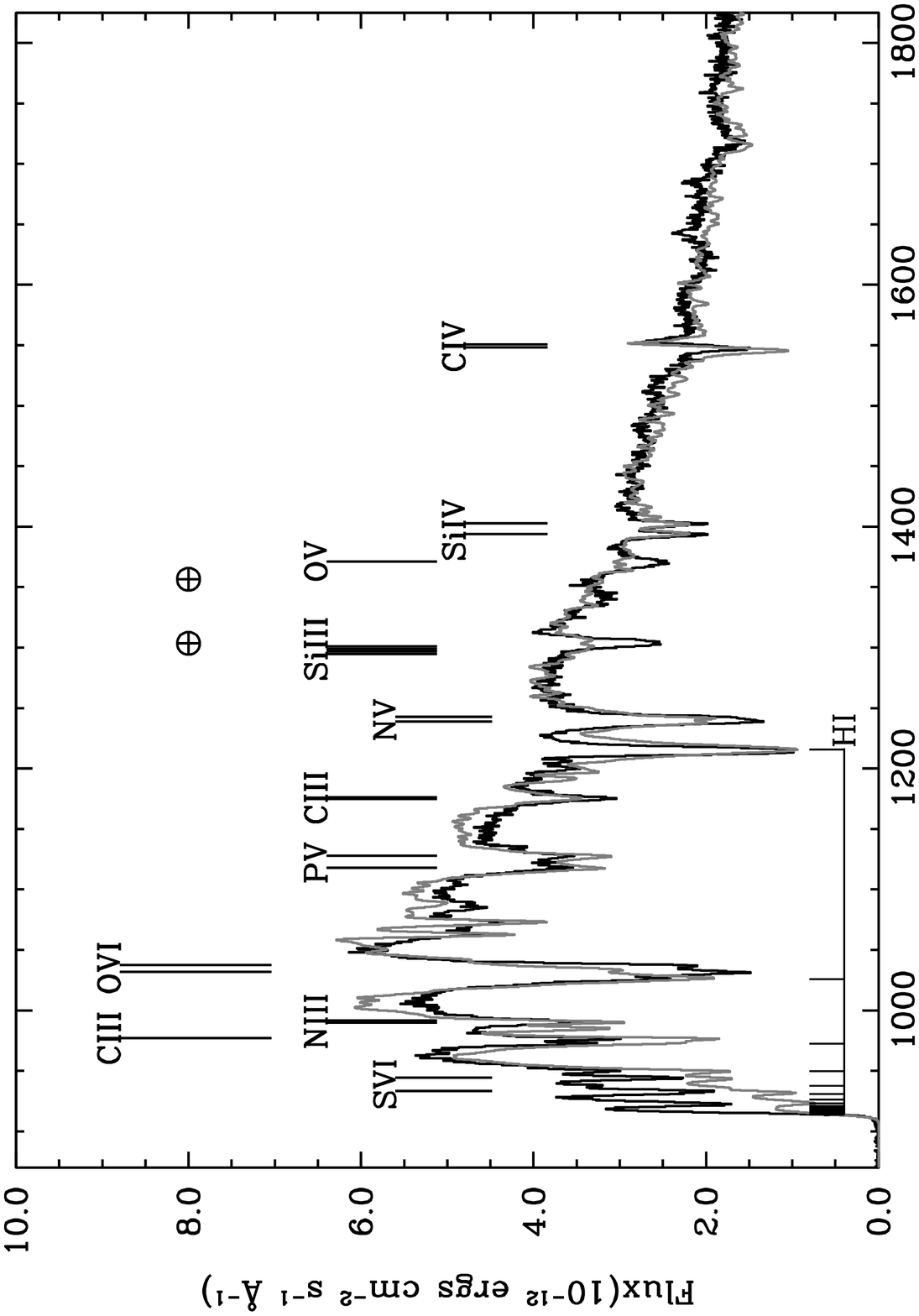}
\end{figure}

\pagebreak
\newpage
\end{document}